\documentclass[prl,superscriptaddress,amsmath,amssymb,aps]{revtex4-1}

\usepackage{graphicx}
\usepackage{dcolumn}
\usepackage{bm}
\usepackage{hyperref}
\usepackage{setspace}  

\hypersetup{
    pdfnewwindow=true,      
    colorlinks=true,        
    linkcolor=blue,         
    citecolor=blue,        
    filecolor=blue,         
    urlcolor=blue           
}

\newcommand{\aap}{Astron. Astrophys.}

\newcommand{\apjl}{Astrophys. J. Lett.}

\newcommand{\mnras}{Mon. Notices Royal Astron. Soc}

\def\spose#1{\hbox to 0pt{#1\hss}}
\def\simlt{\mathrel{\spose{\lower 3pt\hbox{$\mathchar"218$}}
     \raise 2.0pt\hbox{$\mathchar"13C$}}}
\def\simgt{\mathrel{\spose{\lower 3pt\hbox{$\mathchar"218$}}
     \raise 2.0pt\hbox{$\mathchar"13E$}}}

\begin{document}

\preprint{APS/123-QED}

\title{A Candidate Electromagnetic Counterpart to the Binary Black Hole Merger Gravitational Wave Event S190521g\footnote{At the time of writing, LIGO has not yet officially confirmed this event. We still refer to it in this paper using the S-* naming syntax to acknowledge this.}}


\author{M.J. Graham}
\affiliation{Cahill Center for Astronomy $\&$ Astrophysics, California Institute of Technology, 1200 E. California Blvd., Pasadena, CA 91125, USA}
\thanks{mjg@caltech.edu}
\author{K.E.S. Ford}
\affiliation{Department of Science, CUNY-BMCC, 199 Chambers St., New York, NY 10007, USA}
\affiliation{Department of Astrophysics, American Museum of Natural History, Central Park West, New York, NY 10028, USA}
\affiliation{Physics Program, The Graduate Center, CUNY, New York, NY 10016, USA}
\author{B. McKernan}
\affiliation{Department of Science, CUNY-BMCC, 199 Chambers St., New York, NY 10007, USA}
\affiliation{Department of Astrophysics, American Museum of Natural History, Central Park West, New York, NY 10028, USA}
\affiliation{Physics Program, The Graduate Center, CUNY, New York, NY 10016, USA}
\author{N.P. Ross}
\affiliation{Institute for Astronomy, University of Edinburgh, Royal Observatory, Blackford Hill, Edinburgh EH9 3 HJ, UK}
\author{D. Stern}
\affiliation{Jet Propulsion Laboratory, California Institute of Technology, Pasadena, CA 91109, USA}
\author{K. Burdge}
\affiliation{Cahill Center for Astronomy $\&$ Astrophysics, California Institute of Technology, 1200 E. California Blvd., Pasadena, CA 91125, USA}
\author{M. Coughlin}
\affiliation{Division of Physics, Mathematics, and Astronomy, California Institute of Technology, Pasadena, CA 91125, USA}
\affiliation{School of Physics and Astronomy, University of Minnesota, Minneapolis, Minnesota 55455, USA}
\author{S.G. Djorgovski}
\affiliation{Cahill Center for Astronomy $\&$ Astrophysics, California Institute of Technology, 1200 E. California Blvd., Pasadena, CA 91125, USA}
\author{A.J. Drake}
\affiliation{Cahill Center for Astronomy $\&$ Astrophysics, California Institute of Technology, 1200 E. California Blvd., Pasadena, CA 91125, USA}
\author{D. Duev}
\affiliation{Cahill Center for Astronomy $\&$ Astrophysics, California Institute of Technology, 1200 E. California Blvd., Pasadena, CA 91125, USA}
\author{M. Kasliwal}
\affiliation{Cahill Center for Astronomy $\&$ Astrophysics, California Institute of Technology, 1200 E. California Blvd., Pasadena, CA 91125, USA}
\author{A.A. Mahabal}
\affiliation{Cahill Center for Astronomy $\&$ Astrophysics, California Institute of Technology, 1200 E. California Blvd., Pasadena, CA 91125, USA}
\author{S. van Velzen}
\affiliation{Department of Astronomy, University of Maryland, College Park, MD 20742, USA}
\affiliation{Center for Cosmology and Particle Physics, New York University, NY 10003, USA}
\author{J. Belecki}
\affiliation{Caltech Optical Observatories, California Institute of Technology, Pasadena, CA 91125, USA}
\author{E.C. Bellm}
\affiliation{DIRAC Institute, Department of Astronomy, University of Washington, 3910 15th Ave. NE, Seattle, WA 98195, USA}
\author{R. Burruss}
\affiliation{Caltech Optical Observatories, California Institute of Technology, Pasadena, CA 91125, USA}
\author{S.B. Cenko}
\affiliation{Astrophysics Science Division, NASA Goddard Space Flight Center, MC 661, Greenbelt, MD 20771, USA}
\affiliation{Joint Space-Science Institute, University of Maryland, College Park, MD 20742, USA}
\author{V. Cunningham}
\affiliation{Department of Astronomy, University of Maryland, College Park, MD 20742, USA}
\author{G. Helou}
\affiliation{IPAC, California Institute of Technology, 1200 E. California Blvd, Pasadena, CA 91125, USA}
\author{S.R. Kulkarni}
\affiliation{Cahill Center for Astronomy $\&$ Astrophysics, California Institute of Technology, 1200 E. California Blvd., Pasadena, CA 91125, USA}
\author{F.J. Masci}
\affiliation{IPAC, California Institute of Technology, 1200 E. California Blvd., Pasadena, CA 91125, USA}
\author{T. Prince}
\affiliation{Cahill Center for Astronomy $\&$ Astrophysics, California Institute of Technology, 1200 E. California Blvd., Pasadena, CA 91125, USA}
\author{D. Reiley}
\affiliation{Caltech Optical Observatories, California Institute of Technology, Pasadena, CA 91125, USA}
\author{H. Rodriguez}
\affiliation{Caltech Optical Observatories, California Institute of Technology, Pasadena, CA 91125, USA}
\author{B. Rusholme}
\affiliation{IPAC, California Institute of Technology, 1200 E. California Blvd., Pasadena, CA 91125, USA}
\author{R.M. Smith}
\affiliation{Caltech Optical Observatories, California Institute of Technology, Pasadena, CA 91125, USA}
\author{M.T. Soumagnac}
\affiliation{Lawrence Berkeley National Laboratory, 1 Cyclotron Road, Berkeley, CA 94720, USA}
\affiliation{Department of Particle Physics and Astrophysics, Weizmann Institute of Science, Rehovot 76100, Israel}

\date{\today}

\begin{abstract}
We report the first plausible optical electromagnetic (EM) counterpart to a (candidate) binary black hole (BBH) merger. Detected by the Zwicky Transient Facility (ZTF), the EM flare is consistent with expectations for a kicked BBH merger in the accretion disk of an active galactic nucleus (AGN) \citep{McK19a}, and is unlikely ($<O(0.01\%$)) due to intrinsic variability of this source.  The lack of color evolution implies that it is not a supernovae and instead is strongly suggestive of a constant temperature shock. Other false-positive events, such as microlensing or a tidal disruption event, are ruled out or constrained to be $<O(0.1\%$). If the flare is associated with S190521g, we find plausible values of: total mass $ M_{\rm BBH} \sim 100 M_{\odot}$, kick velocity $v_k \sim 200\, {\rm km}\, {\rm s}^{-1}$ at $\theta \sim 60^{\circ}$ in a disk with aspect ratio $H/a \sim 0.01$ (i.e., disk height $H$ at radius $a$) and gas density $\rho \sim 10^{-10}\, {\rm g}\, {\rm cm}^{-3}$. The merger could have occurred at a disk migration trap ($a \sim 700\, r_{g}$; $r_g \equiv G M_{\rm SMBH} / c^2$, where $M_{\rm SMBH}$ is the mass of the AGN supermassive black hole). The combination of parameters implies a significant spin for at least one of the black holes in S190521g. The timing of our spectroscopy prevents useful constraints on broad-line asymmetry due to an off-center flare. We predict a repeat flare in this source due to a re-encountering with the disk in $\sim 1.6\, {\rm yr}\, (M_{\rm SMBH}/10^{8}M_{\odot})\, (a/10^{3}r_{g})^{3/2}$.

\end{abstract}

\maketitle

{\it Introduction.---}  The Laser Interferometer Gravitational wave (GW) Observatory (LIGO) is now detecting binary black hole (BBH) mergers at a high rate in the local ($z<1$) Universe \citep{Abbott19}. The two main channels to BBH mergers are believed to be field binary star evolution \citep[e.g.,][]{Belczynski10, deMink16} and dynamical encounters. Dynamical mergers can occur in globular clusters \citep{Rodriquez16a,Rodriquez16b}, galactic nuclei \citep{Antonini14, AntoniniRasio16,Fragione19}, and in gas disks in galactic nuclei \citep{McK12, McK14, Bellovary16, Bartos17, Stone17, McK18, Secunda19, Yang19, McK19b}. Mergers involving $>50\, M_{\odot}$ black holes (BHs) are unlikely to involve field binary stars \citep{Woosley17}. Rather, massive mergers suggest a dynamical origin, likely in a deep potential where kicked merger products can be retained \citep{Gerosa19}. Several massive mergers may have already been detected, including GW170929 \citep{Chatziioannou19} and GW170817A \citep{NewBin19} (not to be confused with the binary neutron star merger GW170817). A dynamical origin for these mergers implies a much larger number of lower mass mergers from the same channel. Electromagnetic (EM) counterparts are hard to generate in the absence of gas. EM counterparts to supermassive BBH mergers in gas disks are well studied \citep[e.g.][]{Bogdanovic08, Rossi10,Corrales10}, but stellar-origin BBH mergers in active galactic nucleus (AGN) disks can also yield a significant, detectable EM counterpart \citep{McK19a}. 

The Zwicky Transient Facility (ZTF) is a state-of-the-art time-domain survey employing a $47\, {\rm deg}^2$ field-of-view camera on the Palomar 48-inch Samuel Oschin Schmidt telescope \citep{Bellm19,Graham19}. A public survey covers the visible northern sky every three nights in $g$- and $r$-bands to $\sim 20.5$ mag \citep{Bellm19b}. Other observing programs cover smaller areas to greater depth, with higher cadence, or with an additional $i$-band filter. Alerts are generated in real time for all $\geq 5\sigma$ transient detections from difference imaging, and those from the public survey are issued to the community \citep{Patterson19}.

{\it Searching for counterparts.---}
For the 21 LIGO BBH merger triggers in observing run O3a (2019 April 1 - September 30), we identified possible AGN which lay within the 90$\%$ confidence limit region and within the $3\sigma$ limits of the marginal distance distribution integrated over the sky. AGN were identified from the Million Quasar Catalog v6.4 \citep{Flesch19}. Any flare associated with the BBH merger should present within a few days to weeks \citep{McK19a} and so we determined the subset of AGN which were associated with a ZTF alert $\leq 60$ days post-LIGO trigger. Here we present our most promising EM counterpart to a BBH GW event based on a Bayesian changepoint analysis (Graham et al., in preparation).

The event S190521g was observed by both LIGO detectors and the VIRGO detector at 2019-05-21 03:02:29 UTC with a false alarm rate of $3.8 \times 10^{-9}$ Hz (FAR $= 1/8.3$~yr) \citep{gcn24621}. It has a luminosity distance of $3931 \pm 953$ Mpc and was classified as a BBH merger with 97\% certainty. ZTF observed 48\% of the 765 deg$^2$ 90\% localization region of S190521g (half of the localization region is in the southern sky). Alert ZTF19abanrhr (see Fig. 1), first announced $\sim 34$~days after the GW event and associated with AGN J124942.3+344929 at $z=0.438$ (hereafter J1249+3449), was identified as potentially interesting. The AGN is located at the 78\% spatial contour and $1.6\, (0.7)\, \sigma$ from the peak marginal (conditional) luminosity distance. If we convolve the marginal distance distribution for the LIGO event \cite{Singer16} with the quasar luminosity function \cite{Hopkins07} and assume a survey depth of 20.5~mag and a flare probability of $10^{-4}$ per quasar (see below), we would expect to find $10^{-5}$ events in the area and timeframe considered.

From a fit to the H$\beta$ line profile of the AGN, using the QSFit routine \cite{Calderone17}, we find the mass of the central supermassive black hole (SMBH) spans $M_{\rm SMBH} = [1,10] \times 10^8~ M_{\odot}$ and therefore the pre-flare luminosity is $L_{\rm bol}/L_{\rm Edd} = [0.02-0.23]$ relative to the Eddington luminosity. From the ZTF lightcurve, J1249+3449 varied by only a few percent of its mean flux level ($\sim 19.1$ mag in $g$-band) over the $15$ months prior to S190521g. A flare peaking $\sim 50$ days after the GW trigger elevated the flux by $\sim 0.3$~mag (equivalent to $\sim 10^{45}\, {\rm erg}\, {\rm s}^{-1}$) for $\sim 50$ days, assuming a typical quasar bolometric correction factor \citep{Runnoe12}. The total energy released by the flare is therefore $O(10^{51}\, {\rm erg})$. 

\begin{figure}
   \centering  
   \includegraphics[width=0.5\textwidth]{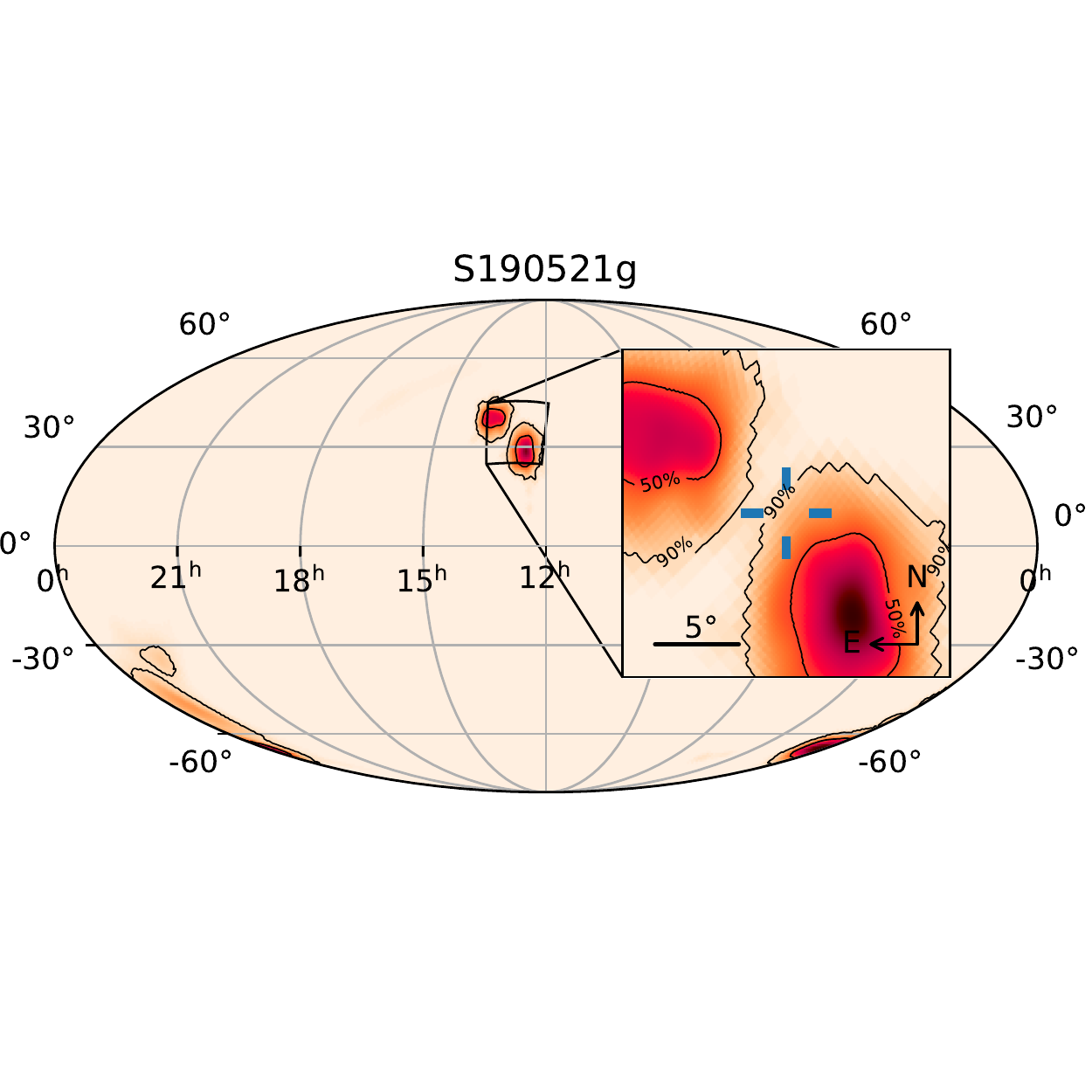}
   \includegraphics[width=0.4\textwidth]{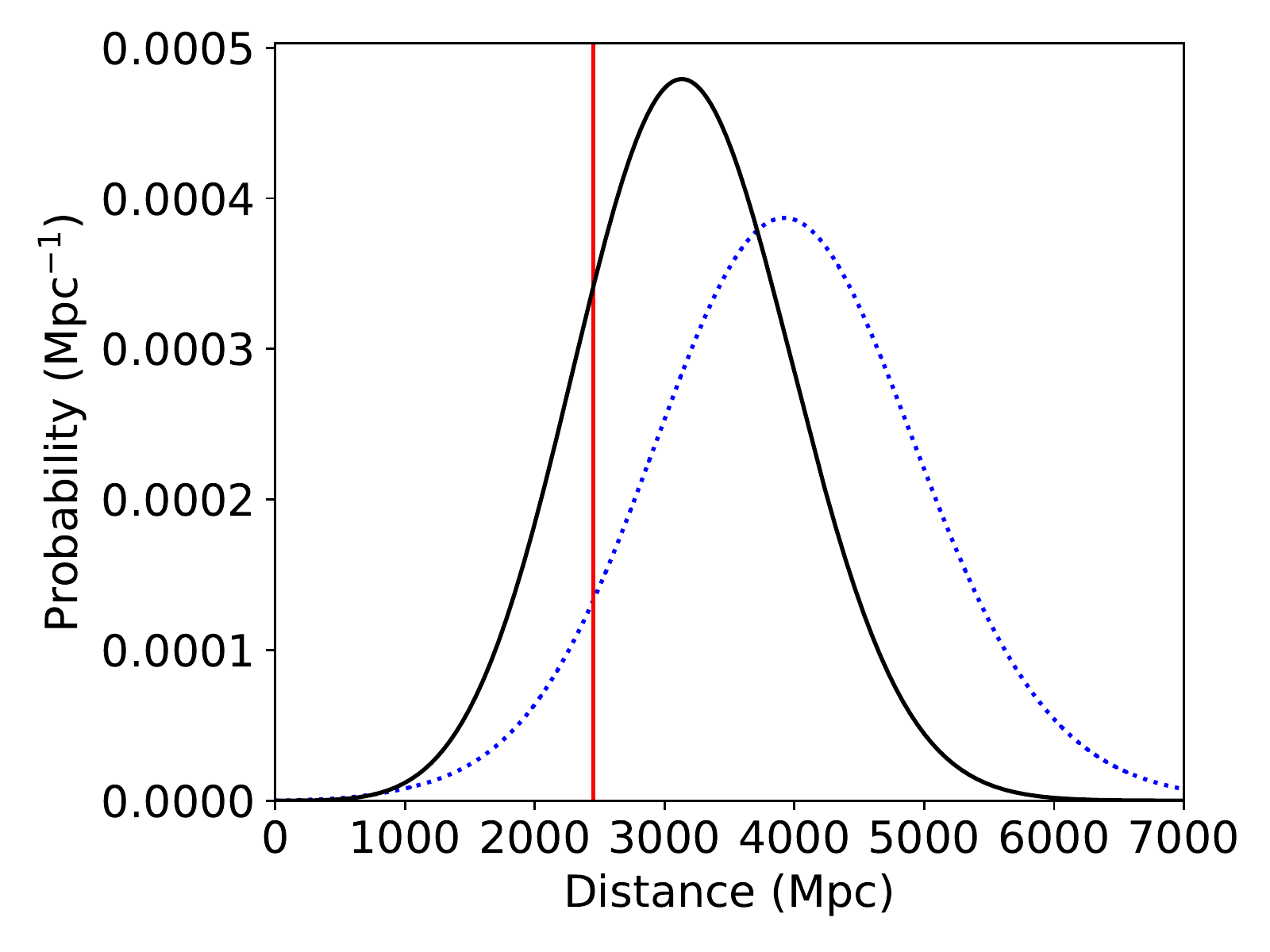}
   \caption{Left panel: A Mollweide projection  of the 50\% and 90\% LIGO localization regions for S190521g (with 44\%/56\% in the northern/southern hemisphere) and the location of ZTF19abanrhr (within the 78\% contour). ZTF covered 48\% of the 90\% region and contours at declination $< -30^\circ$ indicate southern hemisphere regions not covered by ZTF. Right panel: The marginal luminosity distance distribution integrated over the sky (dotted blue line) for S190521g as well as the conditional distance distribution (black line) at the position of ZTF19abanrhr. The red line corresponds to the luminosity distance of ZTF19abanrhr, assuming a Planck15 cosmology \citep{Ade16}.}   
\label{fig:ligo}
\end{figure}

\begin{figure}
   \centering  
   \includegraphics[width=0.5\textwidth]{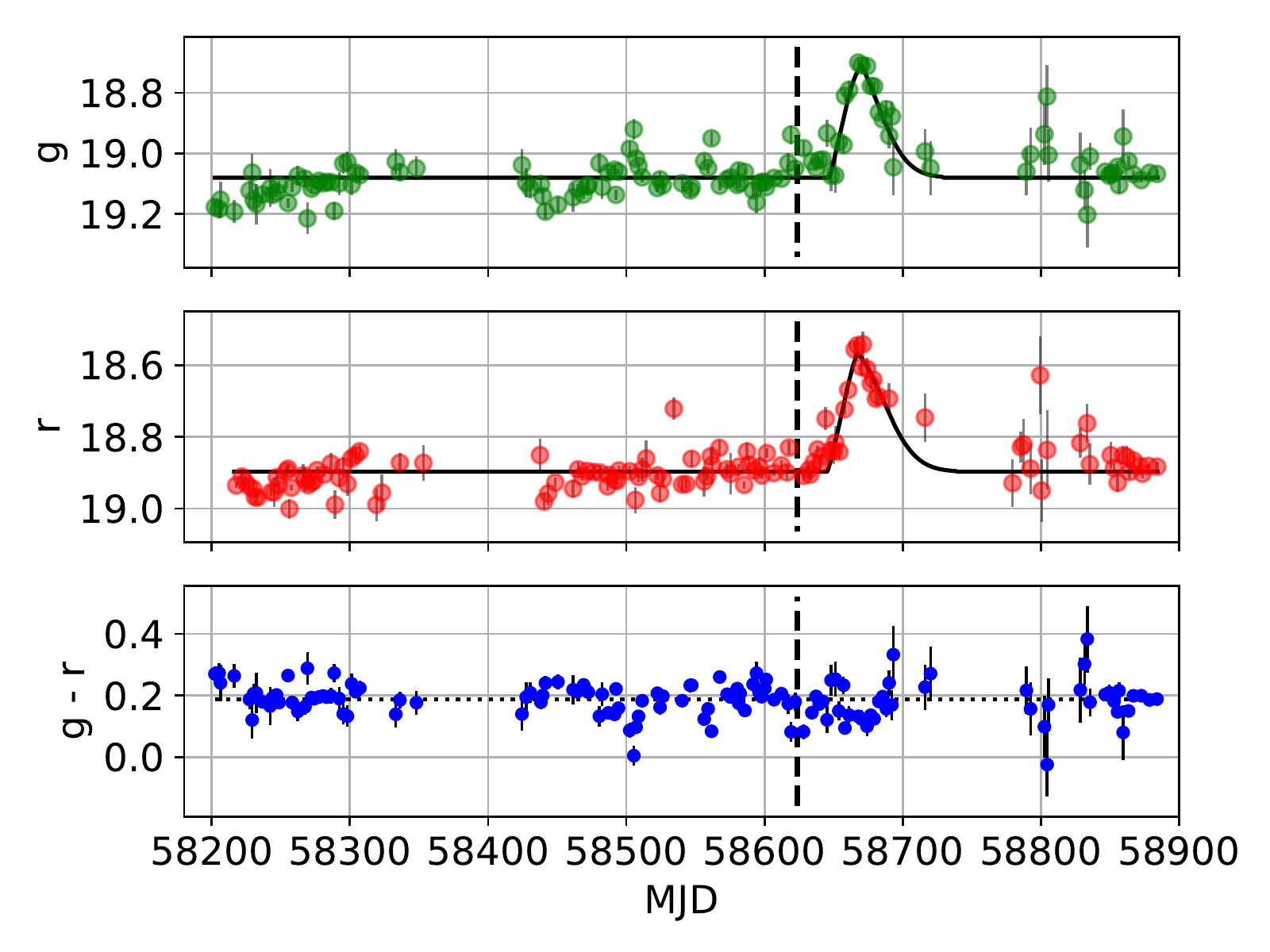}
   \caption{ZTF $g$-band photometry, $r$-band photometry, and $g - r$ color for J1249+3449 over the past 25 months. The flare beginning MJD $\sim$ 58650 represents a $5\sigma$ departure from the ZTF baseline for this source. The flare emission is fit according to the model described in the text and assuming a linear model for the source continuum behaviour over time. The dashed vertical line corresponds to the S190521g trigger time.}   
\label{fig:ztflc}
\end{figure}

\begin{figure}
   \centering  
   \includegraphics[width=0.5\textwidth]{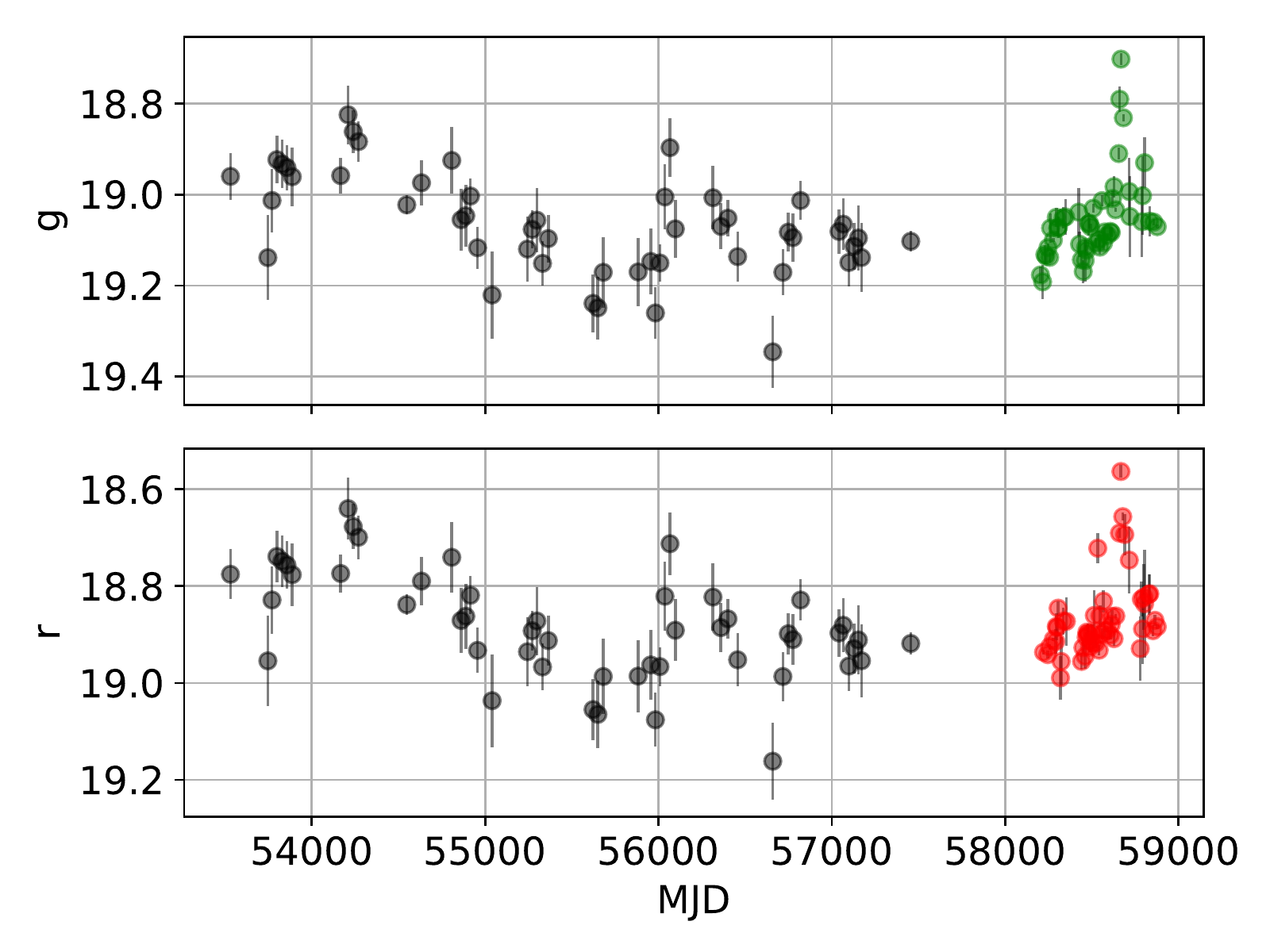}
   \caption{Lightcurve for J1249+3449, including an additional decade of CRTS photometry (binned at 15 day intervals).  ZTF data is binned in 3 day intervals, with $g$- and $r$-band data corrected to the CRTS photometric system using median offsets of 0.52~mag for $g$-band and 0.34~mag for $r$-band.}   
\label{fig:ztfandcrtslc}
\end{figure}

{\it False positives.---}
We consider and rule out, or at least constrain, several possible causes of the ZTF19abanrhr flaring event, such as AGN variability, a supernova, microlensing, and the tidal disruption of a star by an SMBH.

AGN are intrinsically variable, often on quite short timescales \citep{Stern18,Ross18}. However, from Fig.~\ref{fig:ztflc}, this AGN has had a relatively constant luminosity for a year around the flare. We applied models consisting of a generic flare profile (Gaussian rise, exponential decay) superimposed on a linear luminosity model to ZTF lightcurves of all detected sources in the larger {\it WISE}-selected R90 catalogue of 4.5 million high-probability quasar candidates, of which 2.5 million are within the area of sky covered by ZTF and 603,000 are spectroscopically confirmed quasars \citep{Assef18}. We exclude 2912 known blazars and select objects where the flare model is strongly preferred over the linear model (i.e., change in the Bayesian information criterion $\Delta BIC > 10$), the flare is detected in both $g$- and $r$-bands, has at least a 25\% increase in flux, and lasts $\geq 20$ days in the observed frame.  This gives 393 events, of which 209 produced a ZTF alert (the remaining 182 were $< 5 \sigma$ detections above background, and therefore did not produce alerts). 

AGN variability is commonly described statistically as a damped random walk (DRW) process \citep{Kelly09,Moreno19}. If the flare is consistent with this then the same parameterized DRW model (within the confidence limits on the model parameters) should describe the time series with and without the flare \citep{Graham17}. Applying this constraint to both $g$- and $r$-band data reduces the number of flares similar to ZTF19abanrhr (i.e., not attributable to regular AGN activity with greater than $3\sigma$ confidence) to 13. Graham et al. (in preparation) provides more details on the search and the full identified sample. In summary, this analysis shows that the probability of a flare $+$ linear model randomly fitting any given ZTF AGN lightcurve is $\sim 5 \times 10^{-6}$. 

Fig.~\ref{fig:ztfandcrtslc} shows that a decade-long baseline reveals evidence for more significant variability in J1249+3449. Note that these data, from the Catalina Real-time Transient Survey \citep[CRTS;][]{Drake09}, are noisier than ZTF (a result of a 0.7-m survey telescope vs. a 1.2-m survey telescope), and are binned at 15~day intervals for clarity in the plot. Using the DRW model parameters from the CRTS data, which characterize the overall variability of the source, we simulated the observed ZTF light curve 250,000 times and find an equivalent flare (i.e., matching the selection criteria described above) in four instances. The event is thus very unlikely to arise from AGN activity in this particular source (i.e., $\sim O(0.002\%)$. Similarly, to address the look-elsewhere effect, we produced 1000 simulations of the full sample of 3255 AGN in the 90\% three-dimensional localization region of S190521g using their CRTS DRW parameterizations and ZTF time sampling. We find a comparable AGN flare in just five simulations, i.e., $O(0.5\%)$ chance of a false positive, prior to visual inspection.  

Supernovae can occur in AGN \citep[e.g.,][]{Assef18sn}, although the rate is likely small ($>2 \times 10^{-7}\, {\rm AGN}^{-1}\, {\rm yr}^{-1}$ in the {\it WISE} sample). Even with a $O(10^{51}{\rm erg})$ energy output, we expect rise times of $O(20-50)$~days and a decay time or plateau of $\sim 100 - 200$~days \citep{Kasen10}. The flare in Fig.~\ref{fig:ztflc} lasts $40$ days observed-frame, or only $28$ days rest-frame which is a poor match to supernova lightcurves. In addition, supernovae evolve in color over time \citep{Foley11} whereas this flare is uniform with color over time, suggestive of a shock or accretion, rather than a supernova.  We therefore rule out a supernova as a likely false positive.

Microlensing, with an expected rate of $O(10^{-4})$ per AGN \citep{Lawrence16}, is uniform in color at restframe UV/optical bands and is also expected for AGN. However, the expected characteristic timescale for microlensing is $O$(yrs) \citep{Lawrence16}, which is inconsistent with the several week ZTF19abanrhr flare. Assuming a $M_{\odot}$ lens in the source galaxy, we require the lens to orbit at $\sim 1{\rm kpc}$ at $200 {\rm km \, s^{-1}}$ in order to match the timescale ($\sim 2 \times 10^6 {\rm s}$) and magnification ($\sim 1.4$) of this event; assuming a population of $O(10^{10})$ stars in appropriate orbits, geometric considerations produce a rate of $O(10^{-5}) \, {\rm events \, yr^{-1} \, AGN^{-1}}$.


Tidal disruption events (TDEs) also occur in AGN. Stellar disruptions can occur around the central SMBH in a galaxy, but only for $M_{\rm SMBH} \simlt 10^{8} \,M_{\odot}$ \citep[for a non-spinning SMBH;][]{rees88}. TDEs can also occur around small BHs in AGN disks, but as neutron star (NS) or white dwarf (WD) disruptions. EM counterparts to BH-NS tidal disruptions in AGN disks at $z<0.5$ should span $\sim [4,113]\, (f_{\rm AGN}/0.1)\, {\rm yr^{-1}}$ where $f_{\rm AGN}$ is the fraction of BBH mergers expected from the AGN channel \citep{McK20}.  The expected integrated total energy of such events is $O(10^{52}\, {\rm erg})$ \citep{Cannizzaro20}, an order of magnitude more powerful than ZTF19abanrhr. Such an event would also produce a GW signal unlike what was observed based on the inferred chirp mass $M_c$ discussed below for S190521g, and the absence of any other reported LIGO triggers with an appropriate spatial and temporal coincidence). BH-WD disruptions lead to underluminous Type Ia SN with integrated energy $10^{49-51}{\rm erg}$, generally less luminous than ZTF19abanrhr, and decay over a year, and so are ruled out \citep{rosswog09}.

{\it Testing the candidate counterpart.---}
We can derive an approximate mass for any reported GW event from the distance ($d_{L}$) and sky area ($A_{90}$, the 90\% confidence interval for sky area) reported in the public GW event alerts. Specifically, $A_{90} \propto$ SNR$^{-2}$ \citep[e.g.][]{Berry15} and SNR $\propto M_c^{5/6}\, d_{L}^{-1}$ \citep{FinnChernoff93}. Deriving the proportionality constant for a 3-detector system for $A_{90} \propto$ SNR$^{-2}$ from GW190412 \citep{2020arXiv200408342T}, we estimate SNR$\sim8.6$ for S190521g. Assuming equal mass components for this rough calculation, that ZTF19abanrhr is related to S190521g, and using a binary NS range of 110~Mpc (LIGO Hanford) to determine detector sensitivity during the S190521g detection, we estimate a source-frame total mass for $M_{\rm BBH} \sim 150\, M_{\odot}$ (roughly accurate to a factor of 2, ${\rm O}(100\, M_{\odot})$, and plausibly in the upper mass gap).


Gravitational radiation from merging unequal mass BBH carries linear momentum, so the BBH center of mass recoils \citep{Campanelli07,Gonzalez07}. For a BBH merger product kicked with velocity $v_{k}$ in an AGN disk, gravitationally bound gas ($R_{\rm bound} < GM_{\rm BBH}/v_k^{2}$) attempts to follow the BH of mass $M_{\rm BBH}$, but collides with the surrounding disk gas, producing a bright off-center hotspot at UV/optical wavelengths \citep{McK19a}. The radius of gravitationally bound gas is 
\begin{equation}
    \frac{R_{\rm{bound}}}{R_{H}}=0.34 \left(\frac{q}{10^{-6}} \right)^{2/3} \left( \frac{a}{10^{3}\, r_{g}}\right)^{-1} \left(\frac{v_k}{200\, {\rm km}\, {\rm s}^{-1}} \right)^{-2}
\end{equation}
where $R_{H}=a(q/3)^{1/3}$ is the Hill radius of the BH, $a$ is the BH orbit semi-major axis in units of $r_{g} \equiv GM_{\rm SMBH}/c^{2}$, and $q=M_{\rm BBH}/M_{\rm SMBH}$ is the mass ratio of the BBH to the central SMBH. The total energy delivered to the bound gas is $E_{\rm b}=1/2\, M_{\rm b}\, v_{\rm k}^{2}=3/2\, N\, k_{B}\, T_{\rm b}$ where $M_{\rm b} = N\, m_{H}$ is the mass of the bound gas expressed as $N$ atoms of Hydrogen (mass $m_{H}$), $k_B$ is the Boltzmann constant, and $T_{\rm b}$ is the average temperature of the post-shock gas. $E_{\rm b}$ is
    \begin{equation}
      E_{\rm b} = 3\times 10^{45}\, {\rm erg}\left(\frac{\rho}{10^{-10}\, {\rm g}\ {\rm cm}^{-3}}\right)\left(\frac{M_{\rm BBH}}{100\, M_{\odot}}\right)^{3} \left(\frac{v_k}{200\, {\rm km} {\rm s}^{-1}}\right)^{-4}.
   \end{equation}
The dynamical time in the source-frame associated with the ram pressure shock (or the time for the merger remnant to cross the sphere of bound gas) is  $t_{\rm ram} = R_{\rm bound}/v_{\rm k}=GM_{\rm BBH}/v_{\rm k}^{3}$ or
\begin{equation}
    t_{\rm ram} \sim 20\, {\rm day}\left(\frac{M_{\rm BBH}}{100\, M_{\odot}}\right)\left(\frac{v_{\rm k}}{200\, {\rm km}\, {\rm s}^{-1}}\right)^{-3} 
    \label{eq:tram}
\end{equation}
or $\sim 29$~days observed-frame for the same parameterization given the redshift of J1249+3449. The luminosity increase for this process should scale roughly as ${\rm sin}^2\left( \frac{\pi}{2} \frac {t}{t_{\rm ram}}\right)$ until $t>t_{\rm ram}$, when the kicked BH leaves behind the gas that was gravitationally bound at $t=0$. $E_{\rm b}$ is inadequate to explain ZTF19abanrhr, though it induces a delay time ($t_{\rm ram}$) before the dominant luminosity-producing process can begin.

The BH leaves behind bound gas after $t_{\rm ram}$ and enters unperturbed disk gas at $t>t_{\rm ram}$. Nearby gas is accelerated around the BH, producing a shocked Bondi tail \citep[e.g.,][]{Ostriker99, Antoni19, Tejeda19} which both acts as a drag on the BH and accretes onto it. We approximate the Bondi-Hoyle-Lyttleton (BHL) luminosity as $L_{\rm BHL}=\eta \dot{M}_{\rm BHL}c^{2}$ where $\eta$ is the radiative efficiency and
\begin{equation}
    \dot{M}_{\rm BHL}=\frac{4\pi G^{2}M_{\rm BBH}^{2} \rho}{v_{\rm rel}^{3}},
\end{equation}
with $v_{\rm rel}=v_{k}+c_{s}$ and $c_{s}$ is the gas sound speed. As the BH is decelerated, $\dot{M}_{\rm BHL}$ increases. Since $\dot{M}_{\rm BHL}$ is super-Eddington typically, not all of the gas in $\dot{M}_{\rm BHL}$ may end up accreted, but we assume the shock emerges after gas reprocessing with luminosity
\begin{eqnarray}
    L_{\rm BHL}&\approx&2.5 \times 10^{45}{\rm erg}\, {\rm s}^{-1} \left( \frac{\eta}{0.1}\right)\left( \frac{M_{\rm BBH}}{100\, M_{\odot}}\right)^{2} \nonumber \\
    &\times& \left(\frac{v_{k}}{200\, {\rm km}\, {\rm s}^{-1}} \right)^{-3} \left( \frac{\rho}{10^{-10}\, {\rm g}\, {\rm cm^{-3}}}\right)
    \label{eq:bhl}
\end{eqnarray}
where we assume $c_{s} \sim 50\, {\rm km}\, {\rm s}^{-1}$. Bondi drag slows down the kicked BH from initial kinetic energy $1/2 M_{\rm BBH}v_k^{2}$. The drag force is $\dot{M}_{\rm BHL}v_k$ and is equal to $M_{\rm BBH}v_k/t_{\rm dec}$ where $t_{\rm dec}$ is the source-frame deceleration timescale
\begin{eqnarray}
 t_{\rm dec} &=&224\, {\rm yr} \left(\frac{ v_k}{200\, {\rm km}\, {\rm s}^{-1}} \right)^{3}
\left(\frac{\rho}{10^{-10}\, {\rm g}\, {\rm cm}^{-3}}\right)^{-1} 
\nonumber \\
&\times& \left(\frac{M_{\rm BBH}}{100\, M_{\odot}}\right)^{-1}.
\end{eqnarray}
Strong kicks ($v_k > 1000\, {\rm km}\, {\rm s}^{-1}$) are possible under  specific binary arrangements \citep{Zloch15,Healy17}, but as $v_{k} \rightarrow 50\, {\rm km}\, {\rm s}^{-1}$, $t_{\rm dec} \sim 3.5$yr. However, if the event is kicked at an angle $\theta$ to the mid-plane ($\theta=0^{\circ}$ is in the disk mid-plane  and $\theta=90^{\circ}$ is straight up out of the disk), then the EM signature ends when the merged BH exits the disk. The source-frame time for the EM signature to end is
\begin{eqnarray}
t_{\rm end} &\approx& 67\, {\rm day} \left( \frac{v_k}{200\, {\rm km}\, {\rm s}^{-1}}\right)^{-1} \left( \frac{a}{700\, r_{g}}\right)  
\nonumber \\
&\times& \left( \frac{M_{\rm SMBH}}{10^{8}M_{\odot}}\right) \left( \frac{H/a}{0.01}\right)\frac{1}{\sin\left({\theta}/{60^{\circ}} \right)}
  \end{eqnarray}
where $a \sim 700\, r_{g}$ is a plausible migration trap location  \citep{Bellovary16}, and $H/a \sim [10^{-3},0.1]$ is the disk aspect ratio (i.e., disk height $H$ at radius $a$), with $\rho \sim O(10^{-10}){\rm g/cm^{3}}$  appropriate at that radius \citep{Sirko03,TQM05}.



For any EM signature generated below the disk photosphere, the signal will emerge on the photon diffusion timescale ($t_{\rm diff}$) which is
\begin{equation}
t_{\rm diff}=8\, {\rm day} \left(\frac{\tau}{100}\right) \left(\frac{H/a}{0.01}\right) \left(\frac{a}{700\, r_{g}}\right) \left(\frac{M_{\rm SMBH}}{10^{8}M_{\odot}}\right)
\end{equation}
in the source-frame, $\tau$ is the optical depth to the midplane (assumed event location). We can treat photon diffusion from the shocked hot-spot by convolving the shock lightcurve with a Maxwell-Boltzmann distribution with mean time $t_{\rm diff}$. This has the effect of smearing out the actual emergent lightcurve from the disk surface. We plot the resulting flare model fit to the ZTF lightcurve in Fig.~\ref{fig:ztflc}, assuming a linear model for the source continuum. We note that a kicked black hole merger remnant will produce a roughly constant temperature shock, and this is consistent with the lack of color evolution for this flare. If ZTF19abanrhr is not an EM counterpart to S190521g, any flare model must account for this observation.

{\it Parameter estimation.---} 
For either the ram pressure shock or the BHL shock, given even modest optical depth, the shape of the observed lightcurve will be dominated by the Maxwell-Boltzmann distribution. From the EM data we find a best fit $t_{\rm diff} =38^{+2}_{-1}$ day (observed frame) and a $t_{\rm delay}=23^{+1}_{-1}$ day (observed frame). We also find a best fit $t_{\rm end}=80$ day (observed, corresponding to $\sim 57$~day rest-frame). We also find the total energy released in the flare ($\sim 10^{51}\, \rm{erg}$). By inspection, the $g-r$ color implies the temperature of the observed flare is too low to permit strong kicks ($v_k > 1000\, {\rm km}\, {\rm s}^{-1}$), and given the relatively brief duration of the flare ($t_{\rm flare} \sim 40$ day in the observed frame, corresponding to $\sim 28$~day rest-frame), we must assume the event ends due to the merger remnant exiting the disk rather than deceleration. Finding $M_{\rm BBH}=O(100\, M_{\odot})$ from the GW data enables us to make order of magnitude estimates for several system parameters from the EM measurements.

Assuming $M_{\rm BBH} \sim 100\, M_{\odot}$ and $t_{\rm
ram} \sim t_{\rm delay}$, we estimate
$v_k \sim 200\, {\rm km}\, {\rm s}^{-1}$ from eqn.~\ref{eq:tram} (note $v_k \propto M_{\rm BBH}^{1/3}$). The total energy released corresponds to $t_{\rm flare}\, L_{\rm
BHL}$, so $L_{\rm BHL} \sim 10^{45}\, {\rm erg}\, {\rm
s}^{-1}$. Thus, $\rho \sim 10^{-10}\, {\rm g}\, {\rm cm}^{-3}$ from
eqn.~\ref{eq:bhl}, assuming the energy release is dominated by the BHL
shock. With $t_{\rm end} \sim 80$~day ($=v_k H/{\rm sin\theta}$), if we assume the merger
happened near where we would expect a migration trap to occur (i.e.,
$a \sim 700\, r_{g}$), then we find an approximate (but degenerate)
combination of $H/a \sim 0.01$ and $\theta \sim 60^{\circ}$ for
$M_{\rm SMBH} \sim 10^{8}\, M_{\odot}$. $M_{\rm BBH}$
and $v_{k}$ are the best constrained parameters, to factors of $\sim2$. But, since the uncertainty in
$M_{\rm SMBH}$ spans approximately an order of magnitude, the other
parameters estimated above are also uncertain to an order of magnitude.

\begin{figure}
   \centering  
   \includegraphics[width=0.5\textwidth]{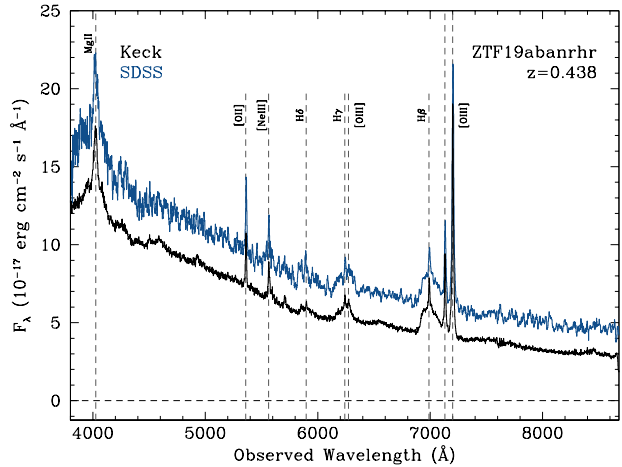}
   \caption{Spectra of J124942.3+344929, the AGN associated with ZTF19abanrhr from SDSS (UT 2006 January 30) and Keck (UT 2020 January 25). Other than fading by $\sim 30\%$, there are no strong spectral changes over the intervening decade (rest-frame).}   
\label{fig:spec}
\end{figure}

{\it Other tests of S190521g association.---}
A kicked BBH merger in an AGN disk will yield an off-center disk flare, producing an asymmetric illumination of the AGN broad line region (BLR) clouds. Depending on the flare luminosity, location, and sightline to the observer, an asymmetric broad line profile will develop within a light-crossing time of the BLR ($R_{\rm BLR}$), and decay over $t_{\rm flare}$ \citep{McK19a}. Unfortunately, the first spectrum of this AGN was taken on UT 2020 January 25, or $\sim 200$ days after the trigger (see Fig.~\ref{fig:spec}). Since the BLR light-crossing time is typically a few weeks, any line broadening effect is no longer present. Therefore we cannot put useful limits on the off-center nature of the flare ZTF19abanrhr.

A modest recoil kick velocity $v_{k}$ corresponds to a small perturbation of the BBH Keplerian orbital velocity $v \sim 10^{4}\, {\rm km}\, {\rm s}^{-1}\, (a/10^{3}r_{g})^{-1/2}$. $v_{k}$ is not large enough to escape the AGN. Therefore, in approximately half an orbital period, the kicked BBH orbit must re-encounter the disk. So, if ZTF19abanrhr is associated with S190521g, we predict a similar flare (driven by Bondi accretion) in this source on a timescale of $1.6\, {\rm yr}\,  (M_{\rm SMBH}/10^{8}M_{\odot})\, (a/10^{3}r_{g})^{3/2}$.

A massive merger in an AGN disk implies a hierarchical origin for at least the primary BH and therefore a high likelihood of significant spin, depending on the merger mass ratio \citep{BertiVolonteri08,GerosaBerti17,Fishbach17,McK19b}. Thus we predict that S190521g includes a significant spin component with the primary BH, and a modest kick velocity \citep{Gerosa16,Varma20}.

{\it Discussion.---}
If we associate ZTF19abanrhr with S190521g, the flare energy is mostly powered by a Bondi accretion tail, which implies a constant color with time, consistent with our data. For a disk thicker than the Hill sphere of the merged BBH, the delay between the GW event and the EM counterpart is $\sim t_{\rm diff}$, the photon diffusion time, which depends on the AGN disk density ($\rho$) and  height ($H$). The temperature measured at the surface of the disk will be lower than the shock temperature, while the rise and decline times will increase, preserving the total energy emitted. 
The strength of this signal (equation 5) depends on the BBH mass squared ($M_{\rm BBH}^{2}$), the recoil kick velocity to the negative three power ($v_{\rm k}^{-3}$), and the AGN disk gas density ($\rho$). So the brightest EM counterparts are for modestly kicked, large mass BBH mergers in dense gas disks.
In anticipation of future small GW error volumes, SMBH mass estimates are needed in as many AGN as possible to constrain the EM follow-up cadence for individual AGN. Other EM generating events will also occur in AGN disks \citep{McK20} and may correspond to peculiar flares observed in several AGN \citep{Graham17,Cannizzaro20}.

{\it Conclusions.---}
We present the first plausible EM counterpart to a BBH merger in an AGN disk. We can rule out most false-positive models at high ($99.9\%$) confidence, and the energetics and color evolution are suggestive of a constant temperature shock, consistent with a kicked BBH merger remnant. We predict a similar repeat flare in this source when the kicked BBH re-encounters the disk on timescale $1.6\, {\rm yr}\,  (M_{\rm SMBH}/10^{8}M_{\odot})\, (a/10^{3}r_{g})^{3/2}$. EM campaigns that trigger follow-up on GW alerts should monitor AGN on multiple cadences, from days to weeks, in order to search for EM counterparts in the AGN channel.

\begin{acknowledgments}
We thank the referees for useful, timely comments that have improved this manuscript. MJG is supported by the NSF grants AST-1518308 and AST-1815034, and the NASA grant 16-ADAP16-0232. KESF \& BM are supported by NSF AST-1831415 and Simons Foundation Grant 533845. KESF \& BM acknowledge extremely useful conversations with Mordecai-Mark MacLow and Pierre Marchand. The work of DS was carried out at the Jet Propulsion Laboratory, California Institute of Technology, under a contract with NASA. MMK acknowledges the GROWTH project funded by the National Science Foundation under Grant No 1545949. MC is supported by NSF PHY-2010970.
Based on observations obtained with the Samuel Oschin Telescope 48-inch and the 60-inch Telescope at the Palomar Observatory as part of the Zwicky Transient Facility project. ZTF is supported by the National Science Foundation under Grant No. AST-1440341 and a collaboration including Caltech, IPAC, the Weizmann Institute for Science, the Oskar Klein Center at Stockholm University, the University of Maryland, the University of Washington, Deutsches Elektronen-Synchrotron and Humboldt University, Los Alamos National Laboratories, the TANGO Consortium of Taiwan, the University of Wisconsin at Milwaukee, and Lawrence Berkeley National Laboratories. Operations are conducted by COO, IPAC, and UW.
The ZTF forced-photometry service was funded under the Heising-Simons Foundation grant 12540303 (PI: Graham).

\end{acknowledgments}

\bibliographystyle{apsrev4-2}
\bibliography{paper}

\begin{thebibliography}{68}%
\makeatletter
\providecommand \@ifxundefined [1]{%
 \@ifx{#1\undefined}
}%
\providecommand \@ifnum [1]{%
 \ifnum #1\expandafter \@firstoftwo
 \else \expandafter \@secondoftwo
 \fi
}%
\providecommand \@ifx [1]{%
 \ifx #1\expandafter \@firstoftwo
 \else \expandafter \@secondoftwo
 \fi
}%
\providecommand \natexlab [1]{#1}%
\providecommand \enquote  [1]{``#1''}%
\providecommand \bibnamefont  [1]{#1}%
\providecommand \bibfnamefont [1]{#1}%
\providecommand \citenamefont [1]{#1}%
\providecommand \href@noop [0]{\@secondoftwo}%
\providecommand \href [0]{\begingroup \@sanitize@url \@href}%
\providecommand \@href[1]{\@@startlink{#1}\@@href}%
\providecommand \@@href[1]{\endgroup#1\@@endlink}%
\providecommand \@sanitize@url [0]{\catcode `\\12\catcode `\$12\catcode
  `\&12\catcode `\#12\catcode `\^12\catcode `\_12\catcode `\%12\relax}%
\providecommand \@@startlink[1]{}%
\providecommand \@@endlink[0]{}%
\providecommand \url  [0]{\begingroup\@sanitize@url \@url }%
\providecommand \@url [1]{\endgroup\@href {#1}{\urlprefix }}%
\providecommand \urlprefix  [0]{URL }%
\providecommand \Eprint [0]{\href }%
\providecommand \doibase [0]{https://doi.org/}%
\providecommand \selectlanguage [0]{\@gobble}%
\providecommand \bibinfo  [0]{\@secondoftwo}%
\providecommand \bibfield  [0]{\@secondoftwo}%
\providecommand \translation [1]{[#1]}%
\providecommand \BibitemOpen [0]{}%
\providecommand \bibitemStop [0]{}%
\providecommand \bibitemNoStop [0]{.\EOS\space}%
\providecommand \EOS [0]{\spacefactor3000\relax}%
\providecommand \BibitemShut  [1]{\csname bibitem#1\endcsname}%
\let\auto@bib@innerbib\@empty
\bibitem [{\citenamefont {{McKernan}}\ \emph {et~al.}(2019)\citenamefont
  {{McKernan}}, \citenamefont {{Ford}}, \citenamefont {{Bartos}} \emph
  {et~al.}}]{McK19a}%
  \BibitemOpen
  \bibfield  {author} {\bibinfo {author} {\bibfnamefont {B.}~\bibnamefont
  {{McKernan}}}, \bibinfo {author} {\bibfnamefont {K.~E.~S.}\ \bibnamefont
  {{Ford}}}, \bibinfo {author} {\bibfnamefont {I.}~\bibnamefont {{Bartos}}},
  \emph {et~al.},\ }\href {https://doi.org/10.3847/2041-8213/ab4886} {\bibfield
   {journal} {\bibinfo  {journal} {\apjl}\ }\textbf {\bibinfo {volume} {884}},\
  \bibinfo {eid} {L50} (\bibinfo {year} {2019})},\ \Eprint
  {https://arxiv.org/abs/1907.03746} {arXiv:1907.03746 [astro-ph.HE]}
  \BibitemShut {NoStop}%
\bibitem [{\citenamefont {{Abbott}}\ \emph {et~al.}(2019)\citenamefont
  {{Abbott}} \emph {et~al.}}]{Abbott19}%
  \BibitemOpen
  \bibfield  {author} {\bibinfo {author} {\bibfnamefont {B.~P.}\ \bibnamefont
  {{Abbott}}} \emph {et~al.},\ }\href
  {https://doi.org/10.1103/PhysRevX.9.031040} {\bibfield  {journal} {\bibinfo
  {journal} {Phys. Rev. X}\ }\textbf {\bibinfo {volume} {9}},\ \bibinfo {pages}
  {031040} (\bibinfo {year} {2019})}\BibitemShut {NoStop}%
\bibitem [{\citenamefont {{Belczynski}}\ \emph {et~al.}(2010)\citenamefont
  {{Belczynski}}, \citenamefont {{Bulik}}, \citenamefont {{Fryer}} \emph
  {et~al.}}]{Belczynski10}%
  \BibitemOpen
  \bibfield  {author} {\bibinfo {author} {\bibfnamefont {K.}~\bibnamefont
  {{Belczynski}}}, \bibinfo {author} {\bibfnamefont {T.}~\bibnamefont
  {{Bulik}}}, \bibinfo {author} {\bibfnamefont {C.~L.}\ \bibnamefont
  {{Fryer}}}, \emph {et~al.},\ }\href
  {https://doi.org/10.1088/0004-637X/714/2/1217} {\bibfield  {journal}
  {\bibinfo  {journal} {\apj}\ }\textbf {\bibinfo {volume} {714}},\ \bibinfo
  {pages} {1217} (\bibinfo {year} {2010})},\ \Eprint
  {https://arxiv.org/abs/0904.2784} {arXiv:0904.2784 [astro-ph.SR]}
  \BibitemShut {NoStop}%
\bibitem [{\citenamefont {{de Mink}}\ and\ \citenamefont
  {{Mandel}}(2016)}]{deMink16}%
  \BibitemOpen
  \bibfield  {author} {\bibinfo {author} {\bibfnamefont {S.~E.}\ \bibnamefont
  {{de Mink}}}\ and\ \bibinfo {author} {\bibfnamefont {I.}~\bibnamefont
  {{Mandel}}},\ }\href {https://doi.org/10.1093/mnras/stw1219} {\bibfield
  {journal} {\bibinfo  {journal} {\mnras}\ }\textbf {\bibinfo {volume} {460}},\
  \bibinfo {pages} {3545} (\bibinfo {year} {2016})},\ \Eprint
  {https://arxiv.org/abs/1603.02291} {arXiv:1603.02291 [astro-ph.HE]}
  \BibitemShut {NoStop}%
\bibitem [{\citenamefont {{Rodriguez}}\ \emph
  {et~al.}(2016{\natexlab{a}})\citenamefont {{Rodriguez}}, \citenamefont
  {{Haster}}, \citenamefont {{Chatterjee}}, \citenamefont {{Kalogera}},\ and\
  \citenamefont {{Rasio}}}]{Rodriquez16a}%
  \BibitemOpen
  \bibfield  {author} {\bibinfo {author} {\bibfnamefont {C.~L.}\ \bibnamefont
  {{Rodriguez}}}, \bibinfo {author} {\bibfnamefont {C.-J.}\ \bibnamefont
  {{Haster}}}, \bibinfo {author} {\bibfnamefont {S.}~\bibnamefont
  {{Chatterjee}}}, \bibinfo {author} {\bibfnamefont {V.}~\bibnamefont
  {{Kalogera}}},\ and\ \bibinfo {author} {\bibfnamefont {F.~A.}\ \bibnamefont
  {{Rasio}}},\ }\href {https://doi.org/10.3847/2041-8205/824/1/L8} {\bibfield
  {journal} {\bibinfo  {journal} {\apjl}\ }\textbf {\bibinfo {volume} {824}},\
  \bibinfo {eid} {L8} (\bibinfo {year} {2016}{\natexlab{a}})},\ \Eprint
  {https://arxiv.org/abs/1604.04254} {arXiv:1604.04254 [astro-ph.HE]}
  \BibitemShut {NoStop}%
\bibitem [{\citenamefont {{Rodriguez}}\ \emph
  {et~al.}(2016{\natexlab{b}})\citenamefont {{Rodriguez}}, \citenamefont
  {{Chatterjee}},\ and\ \citenamefont {{Rasio}}}]{Rodriquez16b}%
  \BibitemOpen
  \bibfield  {author} {\bibinfo {author} {\bibfnamefont {C.~L.}\ \bibnamefont
  {{Rodriguez}}}, \bibinfo {author} {\bibfnamefont {S.}~\bibnamefont
  {{Chatterjee}}},\ and\ \bibinfo {author} {\bibfnamefont {F.~A.}\ \bibnamefont
  {{Rasio}}},\ }\href {https://doi.org/10.1103/PhysRevD.93.084029} {\bibfield
  {journal} {\bibinfo  {journal} {\prd}\ }\textbf {\bibinfo {volume} {93}},\
  \bibinfo {eid} {084029} (\bibinfo {year} {2016}{\natexlab{b}})},\ \Eprint
  {https://arxiv.org/abs/1602.02444} {arXiv:1602.02444 [astro-ph.HE]}
  \BibitemShut {NoStop}%
\bibitem [{\citenamefont {{Antonini}}(2014)}]{Antonini14}%
  \BibitemOpen
  \bibfield  {author} {\bibinfo {author} {\bibfnamefont {F.}~\bibnamefont
  {{Antonini}}},\ }\href {https://doi.org/10.1088/0004-637X/794/2/106}
  {\bibfield  {journal} {\bibinfo  {journal} {\apj}\ }\textbf {\bibinfo
  {volume} {794}},\ \bibinfo {eid} {106} (\bibinfo {year} {2014})},\ \Eprint
  {https://arxiv.org/abs/1402.4865} {arXiv:1402.4865 [astro-ph.GA]}
  \BibitemShut {NoStop}%
\bibitem [{\citenamefont {{Antonini}}\ and\ \citenamefont
  {{Rasio}}(2016)}]{AntoniniRasio16}%
  \BibitemOpen
  \bibfield  {author} {\bibinfo {author} {\bibfnamefont {F.}~\bibnamefont
  {{Antonini}}}\ and\ \bibinfo {author} {\bibfnamefont {F.~A.}\ \bibnamefont
  {{Rasio}}},\ }\href {https://doi.org/10.3847/0004-637X/831/2/187} {\bibfield
  {journal} {\bibinfo  {journal} {\apj}\ }\textbf {\bibinfo {volume} {831}},\
  \bibinfo {eid} {187} (\bibinfo {year} {2016})},\ \Eprint
  {https://arxiv.org/abs/1606.04889} {arXiv:1606.04889 [astro-ph.HE]}
  \BibitemShut {NoStop}%
\bibitem [{\citenamefont {{Fragione}}\ \emph {et~al.}(2019)\citenamefont
  {{Fragione}}, \citenamefont {{Leigh}},\ and\ \citenamefont
  {{Perna}}}]{Fragione19}%
  \BibitemOpen
  \bibfield  {author} {\bibinfo {author} {\bibfnamefont {G.}~\bibnamefont
  {{Fragione}}}, \bibinfo {author} {\bibfnamefont {N.~W.~C.}\ \bibnamefont
  {{Leigh}}},\ and\ \bibinfo {author} {\bibfnamefont {R.}~\bibnamefont
  {{Perna}}},\ }\href {https://doi.org/10.1093/mnras/stz1803} {\bibfield
  {journal} {\bibinfo  {journal} {\mnras}\ }\textbf {\bibinfo {volume} {488}},\
  \bibinfo {pages} {2825} (\bibinfo {year} {2019})},\ \Eprint
  {https://arxiv.org/abs/1903.09160} {arXiv:1903.09160 [astro-ph.GA]}
  \BibitemShut {NoStop}%
\bibitem [{\citenamefont {{McKernan}}\ \emph {et~al.}(2012)\citenamefont
  {{McKernan}}, \citenamefont {{Ford}}, \citenamefont {{Lyra}},\ and\
  \citenamefont {{Perets}}}]{McK12}%
  \BibitemOpen
  \bibfield  {author} {\bibinfo {author} {\bibfnamefont {B.}~\bibnamefont
  {{McKernan}}}, \bibinfo {author} {\bibfnamefont {K.~E.~S.}\ \bibnamefont
  {{Ford}}}, \bibinfo {author} {\bibfnamefont {W.}~\bibnamefont {{Lyra}}},\
  and\ \bibinfo {author} {\bibfnamefont {H.~B.}\ \bibnamefont {{Perets}}},\
  }\href {https://doi.org/10.1111/j.1365-2966.2012.21486.x} {\bibfield
  {journal} {\bibinfo  {journal} {\mnras}\ }\textbf {\bibinfo {volume} {425}},\
  \bibinfo {pages} {460} (\bibinfo {year} {2012})},\ \Eprint
  {https://arxiv.org/abs/1206.2309} {arXiv:1206.2309 [astro-ph.GA]}
  \BibitemShut {NoStop}%
\bibitem [{\citenamefont {{McKernan}}\ \emph {et~al.}(2014)\citenamefont
  {{McKernan}}, \citenamefont {{Ford}}, \citenamefont {{Kocsis}}, \citenamefont
  {{Lyra}},\ and\ \citenamefont {{Winter}}}]{McK14}%
  \BibitemOpen
  \bibfield  {author} {\bibinfo {author} {\bibfnamefont {B.}~\bibnamefont
  {{McKernan}}}, \bibinfo {author} {\bibfnamefont {K.~E.~S.}\ \bibnamefont
  {{Ford}}}, \bibinfo {author} {\bibfnamefont {B.}~\bibnamefont {{Kocsis}}},
  \bibinfo {author} {\bibfnamefont {W.}~\bibnamefont {{Lyra}}},\ and\ \bibinfo
  {author} {\bibfnamefont {L.~M.}\ \bibnamefont {{Winter}}},\ }\href
  {https://doi.org/10.1093/mnras/stu553} {\bibfield  {journal} {\bibinfo
  {journal} {\mnras}\ }\textbf {\bibinfo {volume} {441}},\ \bibinfo {pages}
  {900} (\bibinfo {year} {2014})},\ \Eprint {https://arxiv.org/abs/1403.6433}
  {arXiv:1403.6433 [astro-ph.GA]} \BibitemShut {NoStop}%
\bibitem [{\citenamefont {{Bellovary}}\ \emph {et~al.}(2016)\citenamefont
  {{Bellovary}}, \citenamefont {{Mac Low}}, \citenamefont {{McKernan}},\ and\
  \citenamefont {{Ford}}}]{Bellovary16}%
  \BibitemOpen
  \bibfield  {author} {\bibinfo {author} {\bibfnamefont {J.~M.}\ \bibnamefont
  {{Bellovary}}}, \bibinfo {author} {\bibfnamefont {M.-M.}\ \bibnamefont {{Mac
  Low}}}, \bibinfo {author} {\bibfnamefont {B.}~\bibnamefont {{McKernan}}},\
  and\ \bibinfo {author} {\bibfnamefont {K.~E.~S.}\ \bibnamefont {{Ford}}},\
  }\href {https://doi.org/10.3847/2041-8205/819/2/L17} {\bibfield  {journal}
  {\bibinfo  {journal} {\apjl}\ }\textbf {\bibinfo {volume} {819}},\ \bibinfo
  {eid} {L17} (\bibinfo {year} {2016})},\ \Eprint
  {https://arxiv.org/abs/1511.00005} {arXiv:1511.00005 [astro-ph.GA]}
  \BibitemShut {NoStop}%
\bibitem [{\citenamefont {{Bartos}}\ \emph {et~al.}(2017)\citenamefont
  {{Bartos}}, \citenamefont {{Kocsis}}, \citenamefont {{Haiman}},\ and\
  \citenamefont {{M{\'a}rka}}}]{Bartos17}%
  \BibitemOpen
  \bibfield  {author} {\bibinfo {author} {\bibfnamefont {I.}~\bibnamefont
  {{Bartos}}}, \bibinfo {author} {\bibfnamefont {B.}~\bibnamefont {{Kocsis}}},
  \bibinfo {author} {\bibfnamefont {Z.}~\bibnamefont {{Haiman}}},\ and\
  \bibinfo {author} {\bibfnamefont {S.}~\bibnamefont {{M{\'a}rka}}},\ }\href
  {https://doi.org/10.3847/1538-4357/835/2/165} {\bibfield  {journal} {\bibinfo
   {journal} {\apj}\ }\textbf {\bibinfo {volume} {835}},\ \bibinfo {eid} {165}
  (\bibinfo {year} {2017})},\ \Eprint {https://arxiv.org/abs/1602.03831}
  {arXiv:1602.03831 [astro-ph.HE]} \BibitemShut {NoStop}%
\bibitem [{\citenamefont {{Stone}}\ \emph {et~al.}(2017)\citenamefont
  {{Stone}}, \citenamefont {{Metzger}},\ and\ \citenamefont
  {{Haiman}}}]{Stone17}%
  \BibitemOpen
  \bibfield  {author} {\bibinfo {author} {\bibfnamefont {N.~C.}\ \bibnamefont
  {{Stone}}}, \bibinfo {author} {\bibfnamefont {B.~D.}\ \bibnamefont
  {{Metzger}}},\ and\ \bibinfo {author} {\bibfnamefont {Z.}~\bibnamefont
  {{Haiman}}},\ }\href {https://doi.org/10.1093/mnras/stw2260} {\bibfield
  {journal} {\bibinfo  {journal} {\mnras}\ }\textbf {\bibinfo {volume} {464}},\
  \bibinfo {pages} {946} (\bibinfo {year} {2017})},\ \Eprint
  {https://arxiv.org/abs/1602.04226} {arXiv:1602.04226 [astro-ph.GA]}
  \BibitemShut {NoStop}%
\bibitem [{\citenamefont {{McKernan}}\ \emph {et~al.}(2018)\citenamefont
  {{McKernan}}, \citenamefont {{Ford}}, \citenamefont {{Bellovary}} \emph
  {et~al.}}]{McK18}%
  \BibitemOpen
  \bibfield  {author} {\bibinfo {author} {\bibfnamefont {B.}~\bibnamefont
  {{McKernan}}}, \bibinfo {author} {\bibfnamefont {K.~E.~S.}\ \bibnamefont
  {{Ford}}}, \bibinfo {author} {\bibfnamefont {J.}~\bibnamefont {{Bellovary}}},
  \emph {et~al.},\ }\href {https://doi.org/10.3847/1538-4357/aadae5} {\bibfield
   {journal} {\bibinfo  {journal} {\apj}\ }\textbf {\bibinfo {volume} {866}},\
  \bibinfo {eid} {66} (\bibinfo {year} {2018})},\ \Eprint
  {https://arxiv.org/abs/1702.07818} {arXiv:1702.07818 [astro-ph.HE]}
  \BibitemShut {NoStop}%
\bibitem [{\citenamefont {{Secunda}}\ \emph {et~al.}(2019)\citenamefont
  {{Secunda}}, \citenamefont {{Bellovary}}, \citenamefont {{Mac Low}} \emph
  {et~al.}}]{Secunda19}%
  \BibitemOpen
  \bibfield  {author} {\bibinfo {author} {\bibfnamefont {A.}~\bibnamefont
  {{Secunda}}}, \bibinfo {author} {\bibfnamefont {J.}~\bibnamefont
  {{Bellovary}}}, \bibinfo {author} {\bibfnamefont {M.-M.}\ \bibnamefont {{Mac
  Low}}}, \emph {et~al.},\ }\href {https://doi.org/10.3847/1538-4357/ab20ca}
  {\bibfield  {journal} {\bibinfo  {journal} {\apj}\ }\textbf {\bibinfo
  {volume} {878}},\ \bibinfo {eid} {85} (\bibinfo {year} {2019})},\ \Eprint
  {https://arxiv.org/abs/1807.02859} {arXiv:1807.02859 [astro-ph.HE]}
  \BibitemShut {NoStop}%
\bibitem [{\citenamefont {{Yang}}\ \emph {et~al.}(2019)\citenamefont {{Yang}},
  \citenamefont {{Bartos}}, \citenamefont {{Gayathri}} \emph
  {et~al.}}]{Yang19}%
  \BibitemOpen
  \bibfield  {author} {\bibinfo {author} {\bibfnamefont {Y.}~\bibnamefont
  {{Yang}}}, \bibinfo {author} {\bibfnamefont {I.}~\bibnamefont {{Bartos}}},
  \bibinfo {author} {\bibfnamefont {V.}~\bibnamefont {{Gayathri}}}, \emph
  {et~al.},\ }\href {https://doi.org/10.1103/PhysRevLett.123.181101} {\bibfield
   {journal} {\bibinfo  {journal} {\prl}\ }\textbf {\bibinfo {volume} {123}},\
  \bibinfo {eid} {181101} (\bibinfo {year} {2019})},\ \Eprint
  {https://arxiv.org/abs/1906.09281} {arXiv:1906.09281 [astro-ph.HE]}
  \BibitemShut {NoStop}%
\bibitem [{\citenamefont {{McKernan}}\ \emph
  {et~al.}(2020{\natexlab{a}})\citenamefont {{McKernan}}, \citenamefont
  {{Ford}}, \citenamefont {{O'Shaugnessy}},\ and\ \citenamefont
  {{Wysocki}}}]{McK19b}%
  \BibitemOpen
  \bibfield  {author} {\bibinfo {author} {\bibfnamefont {B.}~\bibnamefont
  {{McKernan}}}, \bibinfo {author} {\bibfnamefont {K.~E.~S.}\ \bibnamefont
  {{Ford}}}, \bibinfo {author} {\bibfnamefont {R.}~\bibnamefont
  {{O'Shaugnessy}}},\ and\ \bibinfo {author} {\bibfnamefont {D.}~\bibnamefont
  {{Wysocki}}},\ }\href {https://doi.org/10.1093/mnras/staa740} {\bibfield
  {journal} {\bibinfo  {journal} {\mnras}\ }\textbf {\bibinfo {volume} {494}},\
  \bibinfo {pages} {1203} (\bibinfo {year} {2020}{\natexlab{a}})},\ \Eprint
  {https://arxiv.org/abs/1907.04356} {arXiv:1907.04356 [astro-ph.HE]}
  \BibitemShut {NoStop}%
\bibitem [{\citenamefont {{Woosley}}(2017)}]{Woosley17}%
  \BibitemOpen
  \bibfield  {author} {\bibinfo {author} {\bibfnamefont {S.~E.}\ \bibnamefont
  {{Woosley}}},\ }\href {https://doi.org/10.3847/1538-4357/836/2/244}
  {\bibfield  {journal} {\bibinfo  {journal} {\apj}\ }\textbf {\bibinfo
  {volume} {836}},\ \bibinfo {eid} {244} (\bibinfo {year} {2017})},\ \Eprint
  {https://arxiv.org/abs/1608.08939} {arXiv:1608.08939 [astro-ph.HE]}
  \BibitemShut {NoStop}%
\bibitem [{\citenamefont {{Gerosa}}\ and\ \citenamefont
  {{Berti}}(2019)}]{Gerosa19}%
  \BibitemOpen
  \bibfield  {author} {\bibinfo {author} {\bibfnamefont {D.}~\bibnamefont
  {{Gerosa}}}\ and\ \bibinfo {author} {\bibfnamefont {E.}~\bibnamefont
  {{Berti}}},\ }\href {https://doi.org/10.1103/PhysRevD.100.041301} {\bibfield
  {journal} {\bibinfo  {journal} {\prd}\ }\textbf {\bibinfo {volume} {100}},\
  \bibinfo {eid} {041301} (\bibinfo {year} {2019})},\ \Eprint
  {https://arxiv.org/abs/1906.05295} {arXiv:1906.05295 [astro-ph.HE]}
  \BibitemShut {NoStop}%
\bibitem [{\citenamefont {{Chatziioannou}}\ \emph {et~al.}(2019)\citenamefont
  {{Chatziioannou}}, \citenamefont {{Cotesta}}, \citenamefont {{Ghonge}} \emph
  {et~al.}}]{Chatziioannou19}%
  \BibitemOpen
  \bibfield  {author} {\bibinfo {author} {\bibfnamefont {K.}~\bibnamefont
  {{Chatziioannou}}}, \bibinfo {author} {\bibfnamefont {R.}~\bibnamefont
  {{Cotesta}}}, \bibinfo {author} {\bibfnamefont {S.}~\bibnamefont {{Ghonge}}},
  \emph {et~al.},\ }\href {https://doi.org/10.1103/PhysRevD.100.104015}
  {\bibfield  {journal} {\bibinfo  {journal} {\prd}\ }\textbf {\bibinfo
  {volume} {100}},\ \bibinfo {eid} {104015} (\bibinfo {year} {2019})},\ \Eprint
  {https://arxiv.org/abs/1903.06742} {arXiv:1903.06742 [gr-qc]} \BibitemShut
  {NoStop}%
\bibitem [{\citenamefont {{Zackay}}\ \emph {et~al.}(2019)\citenamefont
  {{Zackay}}, \citenamefont {{Dai}}, \citenamefont {{Venumadhav}},
  \citenamefont {{Roulet}},\ and\ \citenamefont {{Zaldarriaga}}}]{NewBin19}%
  \BibitemOpen
  \bibfield  {author} {\bibinfo {author} {\bibfnamefont {B.}~\bibnamefont
  {{Zackay}}}, \bibinfo {author} {\bibfnamefont {L.}~\bibnamefont {{Dai}}},
  \bibinfo {author} {\bibfnamefont {T.}~\bibnamefont {{Venumadhav}}}, \bibinfo
  {author} {\bibfnamefont {J.}~\bibnamefont {{Roulet}}},\ and\ \bibinfo
  {author} {\bibfnamefont {M.}~\bibnamefont {{Zaldarriaga}}},\ }\href@noop {}
  {\bibfield  {journal} {\bibinfo  {journal} {arXiv e-prints}\ ,\ \bibinfo
  {eid} {arXiv:1910.09528}} (\bibinfo {year} {2019})},\ \Eprint
  {https://arxiv.org/abs/1910.09528} {arXiv:1910.09528 [astro-ph.HE]}
  \BibitemShut {NoStop}%
\bibitem [{\citenamefont {{Bogdanovi{\'c}}}\ \emph {et~al.}(2008)\citenamefont
  {{Bogdanovi{\'c}}}, \citenamefont {{Smith}}, \citenamefont {{Sigurdsson}},\
  and\ \citenamefont {{Eracleous}}}]{Bogdanovic08}%
  \BibitemOpen
  \bibfield  {author} {\bibinfo {author} {\bibfnamefont {T.}~\bibnamefont
  {{Bogdanovi{\'c}}}}, \bibinfo {author} {\bibfnamefont {B.~D.}\ \bibnamefont
  {{Smith}}}, \bibinfo {author} {\bibfnamefont {S.}~\bibnamefont
  {{Sigurdsson}}},\ and\ \bibinfo {author} {\bibfnamefont {M.}~\bibnamefont
  {{Eracleous}}},\ }\href {https://doi.org/10.1086/521828} {\bibfield
  {journal} {\bibinfo  {journal} {ApJS}\ }\textbf {\bibinfo {volume} {174}},\
  \bibinfo {pages} {455} (\bibinfo {year} {2008})},\ \Eprint
  {https://arxiv.org/abs/0708.0414} {arXiv:0708.0414 [astro-ph]} \BibitemShut
  {NoStop}%
\bibitem [{\citenamefont {{Rossi}}\ \emph {et~al.}(2010)\citenamefont
  {{Rossi}}, \citenamefont {{Lodato}}, \citenamefont {{Armitage}},
  \citenamefont {{Pringle}},\ and\ \citenamefont {{King}}}]{Rossi10}%
  \BibitemOpen
  \bibfield  {author} {\bibinfo {author} {\bibfnamefont {E.~M.}\ \bibnamefont
  {{Rossi}}}, \bibinfo {author} {\bibfnamefont {G.}~\bibnamefont {{Lodato}}},
  \bibinfo {author} {\bibfnamefont {P.~J.}\ \bibnamefont {{Armitage}}},
  \bibinfo {author} {\bibfnamefont {J.~E.}\ \bibnamefont {{Pringle}}},\ and\
  \bibinfo {author} {\bibfnamefont {A.~R.}\ \bibnamefont {{King}}},\ }\href
  {https://doi.org/10.1111/j.1365-2966.2009.15802.x} {\bibfield  {journal}
  {\bibinfo  {journal} {\mnras}\ }\textbf {\bibinfo {volume} {401}},\ \bibinfo
  {pages} {2021} (\bibinfo {year} {2010})},\ \Eprint
  {https://arxiv.org/abs/0910.0002} {arXiv:0910.0002 [astro-ph.HE]}
  \BibitemShut {NoStop}%
\bibitem [{\citenamefont {{Corrales}}\ \emph {et~al.}(2010)\citenamefont
  {{Corrales}}, \citenamefont {{Haiman}},\ and\ \citenamefont
  {{MacFadyen}}}]{Corrales10}%
  \BibitemOpen
  \bibfield  {author} {\bibinfo {author} {\bibfnamefont {L.~R.}\ \bibnamefont
  {{Corrales}}}, \bibinfo {author} {\bibfnamefont {Z.}~\bibnamefont
  {{Haiman}}},\ and\ \bibinfo {author} {\bibfnamefont {A.}~\bibnamefont
  {{MacFadyen}}},\ }\href {https://doi.org/10.1111/j.1365-2966.2010.16324.x}
  {\bibfield  {journal} {\bibinfo  {journal} {\mnras}\ }\textbf {\bibinfo
  {volume} {404}},\ \bibinfo {pages} {947} (\bibinfo {year} {2010})},\ \Eprint
  {https://arxiv.org/abs/0910.0014} {arXiv:0910.0014 [astro-ph.HE]}
  \BibitemShut {NoStop}%
\bibitem [{\citenamefont {{Bellm}}\ \emph
  {et~al.}(2019{\natexlab{a}})\citenamefont {{Bellm}}, \citenamefont
  {{Kulkarni}}, \citenamefont {{Graham}} \emph {et~al.}}]{Bellm19}%
  \BibitemOpen
  \bibfield  {author} {\bibinfo {author} {\bibfnamefont {E.~C.}\ \bibnamefont
  {{Bellm}}}, \bibinfo {author} {\bibfnamefont {S.~R.}\ \bibnamefont
  {{Kulkarni}}}, \bibinfo {author} {\bibfnamefont {M.~J.}\ \bibnamefont
  {{Graham}}}, \emph {et~al.},\ }\href
  {https://doi.org/10.1088/1538-3873/aaecbe} {\bibfield  {journal} {\bibinfo
  {journal} {PASP}\ }\textbf {\bibinfo {volume} {131}},\ \bibinfo {pages}
  {018002} (\bibinfo {year} {2019}{\natexlab{a}})},\ \Eprint
  {https://arxiv.org/abs/1902.01932} {arXiv:1902.01932 [astro-ph.IM]}
  \BibitemShut {NoStop}%
\bibitem [{\citenamefont {{Graham}}\ \emph {et~al.}(2019)\citenamefont
  {{Graham}}, \citenamefont {{Kulkarni}}, \citenamefont {{Bellm}} \emph
  {et~al.}}]{Graham19}%
  \BibitemOpen
  \bibfield  {author} {\bibinfo {author} {\bibfnamefont {M.~J.}\ \bibnamefont
  {{Graham}}}, \bibinfo {author} {\bibfnamefont {S.~R.}\ \bibnamefont
  {{Kulkarni}}}, \bibinfo {author} {\bibfnamefont {E.~C.}\ \bibnamefont
  {{Bellm}}}, \emph {et~al.},\ }\href
  {https://doi.org/10.1088/1538-3873/ab006c} {\bibfield  {journal} {\bibinfo
  {journal} {PASP}\ }\textbf {\bibinfo {volume} {131}},\ \bibinfo {pages}
  {078001} (\bibinfo {year} {2019})},\ \Eprint
  {https://arxiv.org/abs/1902.01945} {arXiv:1902.01945 [astro-ph.IM]}
  \BibitemShut {NoStop}%
\bibitem [{\citenamefont {{Bellm}}\ \emph
  {et~al.}(2019{\natexlab{b}})\citenamefont {{Bellm}}, \citenamefont
  {{Kulkarni}}, \citenamefont {{Barlow}}, \citenamefont {{Feindt}},
  \citenamefont {{Graham}}, \citenamefont {{Goobar}}, \citenamefont {{Kupfer}},
  \citenamefont {{Ngeow}}, \citenamefont {{Nugent}}, \citenamefont {{Ofek}},
  \citenamefont {{Prince}}, \citenamefont {{Riddle}}, \citenamefont
  {{Walters}},\ and\ \citenamefont {{Ye}}}]{Bellm19b}%
  \BibitemOpen
  \bibfield  {author} {\bibinfo {author} {\bibfnamefont {E.~C.}\ \bibnamefont
  {{Bellm}}}, \bibinfo {author} {\bibfnamefont {S.~R.}\ \bibnamefont
  {{Kulkarni}}}, \bibinfo {author} {\bibfnamefont {T.}~\bibnamefont
  {{Barlow}}}, \bibinfo {author} {\bibfnamefont {U.}~\bibnamefont {{Feindt}}},
  \bibinfo {author} {\bibfnamefont {M.~J.}\ \bibnamefont {{Graham}}}, \bibinfo
  {author} {\bibfnamefont {A.}~\bibnamefont {{Goobar}}}, \bibinfo {author}
  {\bibfnamefont {T.}~\bibnamefont {{Kupfer}}}, \bibinfo {author}
  {\bibfnamefont {C.-C.}\ \bibnamefont {{Ngeow}}}, \bibinfo {author}
  {\bibfnamefont {P.}~\bibnamefont {{Nugent}}}, \bibinfo {author}
  {\bibfnamefont {E.}~\bibnamefont {{Ofek}}}, \bibinfo {author} {\bibfnamefont
  {T.~A.}\ \bibnamefont {{Prince}}}, \bibinfo {author} {\bibfnamefont
  {R.}~\bibnamefont {{Riddle}}}, \bibinfo {author} {\bibfnamefont
  {R.}~\bibnamefont {{Walters}}},\ and\ \bibinfo {author} {\bibfnamefont
  {Q.-Z.}\ \bibnamefont {{Ye}}},\ }\href
  {https://doi.org/10.1088/1538-3873/ab0c2a} {\bibfield  {journal} {\bibinfo
  {journal} {PASP}\ }\textbf {\bibinfo {volume} {131}},\ \bibinfo {pages}
  {068003} (\bibinfo {year} {2019}{\natexlab{b}})},\ \Eprint
  {https://arxiv.org/abs/1905.02209} {arXiv:1905.02209 [astro-ph.IM]}
  \BibitemShut {NoStop}%
\bibitem [{\citenamefont {{Patterson}}\ \emph {et~al.}(2019)\citenamefont
  {{Patterson}}, \citenamefont {{Bellm}}, \citenamefont {{Rusholme}},
  \citenamefont {{Masci}}, \citenamefont {{Juric}}, \citenamefont {{Krughoff}},
  \citenamefont {{Golkhou}}, \citenamefont {{Graham}}, \citenamefont
  {{Kulkarni}}, \citenamefont {{Helou}},\ and\ \citenamefont {{Zwicky Transient
  Facility Collaboration}}}]{Patterson19}%
  \BibitemOpen
  \bibfield  {author} {\bibinfo {author} {\bibfnamefont {M.~T.}\ \bibnamefont
  {{Patterson}}}, \bibinfo {author} {\bibfnamefont {E.~C.}\ \bibnamefont
  {{Bellm}}}, \bibinfo {author} {\bibfnamefont {B.}~\bibnamefont {{Rusholme}}},
  \bibinfo {author} {\bibfnamefont {F.~J.}\ \bibnamefont {{Masci}}}, \bibinfo
  {author} {\bibfnamefont {M.}~\bibnamefont {{Juric}}}, \bibinfo {author}
  {\bibfnamefont {K.~S.}\ \bibnamefont {{Krughoff}}}, \bibinfo {author}
  {\bibfnamefont {V.~Z.}\ \bibnamefont {{Golkhou}}}, \bibinfo {author}
  {\bibfnamefont {M.~J.}\ \bibnamefont {{Graham}}}, \bibinfo {author}
  {\bibfnamefont {S.~R.}\ \bibnamefont {{Kulkarni}}}, \bibinfo {author}
  {\bibfnamefont {G.}~\bibnamefont {{Helou}}},\ and\ \bibinfo {author}
  {\bibnamefont {{Zwicky Transient Facility Collaboration}}},\ }\href
  {https://doi.org/10.1088/1538-3873/aae904} {\bibfield  {journal} {\bibinfo
  {journal} {PASP}\ }\textbf {\bibinfo {volume} {131}},\ \bibinfo {pages}
  {018001} (\bibinfo {year} {2019})},\ \Eprint
  {https://arxiv.org/abs/1902.02227} {arXiv:1902.02227 [astro-ph.IM]}
  \BibitemShut {NoStop}%
\bibitem [{\citenamefont {{Flesch}}(2019)}]{Flesch19}%
  \BibitemOpen
  \bibfield  {author} {\bibinfo {author} {\bibfnamefont {E.~W.}\ \bibnamefont
  {{Flesch}}},\ }\href@noop {} {\bibfield  {journal} {\bibinfo  {journal}
  {arXiv e-prints}\ ,\ \bibinfo {eid} {arXiv:1912.05614}} (\bibinfo {year}
  {2019})},\ \Eprint {https://arxiv.org/abs/1912.05614} {arXiv:1912.05614
  [astro-ph.GA]} \BibitemShut {NoStop}%
\bibitem [{\citenamefont {{LIGO/VIRGO Collaboration}}\ \emph
  {et~al.}(2019)\citenamefont {{LIGO/VIRGO Collaboration}} \emph
  {et~al.}}]{gcn24621}%
  \BibitemOpen
  \bibfield  {author} {\bibinfo {author} {\bibnamefont {{LIGO/VIRGO
  Collaboration}}} \emph {et~al.},\ }\href@noop {} {\bibfield  {journal}
  {\bibinfo  {journal} {GCN Circular}\ }\textbf {\bibinfo {volume} {24621}}
  (\bibinfo {year} {2019})}\BibitemShut {NoStop}%
\bibitem [{\citenamefont {{Singer}}\ \emph {et~al.}(2016)\citenamefont
  {{Singer}}, \citenamefont {{Chen}}, \citenamefont {{Holz}}, \citenamefont
  {{Farr}}, \citenamefont {{Price}}, \citenamefont {{Raymond}}, \citenamefont
  {{Cenko}}, \citenamefont {{Gehrels}}, \citenamefont {{Cannizzo}},
  \citenamefont {{Kasliwal}}, \citenamefont {{Nissanke}}, \citenamefont
  {{Coughlin}}, \citenamefont {{Farr}}, \citenamefont {{Urban}}, \citenamefont
  {{Vitale}}, \citenamefont {{Veitch}}, \citenamefont {{Graff}}, \citenamefont
  {{Berry}}, \citenamefont {{Mohapatra}},\ and\ \citenamefont
  {{Mandel}}}]{Singer16}%
  \BibitemOpen
  \bibfield  {author} {\bibinfo {author} {\bibfnamefont {L.~P.}\ \bibnamefont
  {{Singer}}}, \bibinfo {author} {\bibfnamefont {H.-Y.}\ \bibnamefont
  {{Chen}}}, \bibinfo {author} {\bibfnamefont {D.~E.}\ \bibnamefont {{Holz}}},
  \bibinfo {author} {\bibfnamefont {W.~M.}\ \bibnamefont {{Farr}}}, \bibinfo
  {author} {\bibfnamefont {L.~R.}\ \bibnamefont {{Price}}}, \bibinfo {author}
  {\bibfnamefont {V.}~\bibnamefont {{Raymond}}}, \bibinfo {author}
  {\bibfnamefont {S.~B.}\ \bibnamefont {{Cenko}}}, \bibinfo {author}
  {\bibfnamefont {N.}~\bibnamefont {{Gehrels}}}, \bibinfo {author}
  {\bibfnamefont {J.}~\bibnamefont {{Cannizzo}}}, \bibinfo {author}
  {\bibfnamefont {M.~M.}\ \bibnamefont {{Kasliwal}}}, \bibinfo {author}
  {\bibfnamefont {S.}~\bibnamefont {{Nissanke}}}, \bibinfo {author}
  {\bibfnamefont {M.}~\bibnamefont {{Coughlin}}}, \bibinfo {author}
  {\bibfnamefont {B.}~\bibnamefont {{Farr}}}, \bibinfo {author} {\bibfnamefont
  {A.~L.}\ \bibnamefont {{Urban}}}, \bibinfo {author} {\bibfnamefont
  {S.}~\bibnamefont {{Vitale}}}, \bibinfo {author} {\bibfnamefont
  {J.}~\bibnamefont {{Veitch}}}, \bibinfo {author} {\bibfnamefont
  {P.}~\bibnamefont {{Graff}}}, \bibinfo {author} {\bibfnamefont {C.~P.~L.}\
  \bibnamefont {{Berry}}}, \bibinfo {author} {\bibfnamefont {S.}~\bibnamefont
  {{Mohapatra}}},\ and\ \bibinfo {author} {\bibfnamefont {I.}~\bibnamefont
  {{Mandel}}},\ }\href {https://doi.org/10.3847/0067-0049/226/1/10} {\bibfield
  {journal} {\bibinfo  {journal} {ApJS}\ }\textbf {\bibinfo {volume} {226}},\
  \bibinfo {eid} {10} (\bibinfo {year} {2016})},\ \Eprint
  {https://arxiv.org/abs/1605.04242} {arXiv:1605.04242 [astro-ph.IM]}
  \BibitemShut {NoStop}%
\bibitem [{\citenamefont {{Hopkins}}\ \emph {et~al.}(2007)\citenamefont
  {{Hopkins}}, \citenamefont {{Richards}},\ and\ \citenamefont
  {{Hernquist}}}]{Hopkins07}%
  \BibitemOpen
  \bibfield  {author} {\bibinfo {author} {\bibfnamefont {P.~F.}\ \bibnamefont
  {{Hopkins}}}, \bibinfo {author} {\bibfnamefont {G.~T.}\ \bibnamefont
  {{Richards}}},\ and\ \bibinfo {author} {\bibfnamefont {L.}~\bibnamefont
  {{Hernquist}}},\ }\href {https://doi.org/10.1086/509629} {\bibfield
  {journal} {\bibinfo  {journal} {\apj}\ }\textbf {\bibinfo {volume} {654}},\
  \bibinfo {pages} {731} (\bibinfo {year} {2007})},\ \Eprint
  {https://arxiv.org/abs/astro-ph/0605678} {arXiv:astro-ph/0605678 [astro-ph]}
  \BibitemShut {NoStop}%
\bibitem [{\citenamefont {{Calderone}}\ \emph {et~al.}(2017)\citenamefont
  {{Calderone}}, \citenamefont {{Nicastro}}, \citenamefont {{Ghisellini}},
  \citenamefont {{Dotti}}, \citenamefont {{Sbarrato}}, \citenamefont
  {{Shankar}},\ and\ \citenamefont {{Colpi}}}]{Calderone17}%
  \BibitemOpen
  \bibfield  {author} {\bibinfo {author} {\bibfnamefont {G.}~\bibnamefont
  {{Calderone}}}, \bibinfo {author} {\bibfnamefont {L.}~\bibnamefont
  {{Nicastro}}}, \bibinfo {author} {\bibfnamefont {G.}~\bibnamefont
  {{Ghisellini}}}, \bibinfo {author} {\bibfnamefont {M.}~\bibnamefont
  {{Dotti}}}, \bibinfo {author} {\bibfnamefont {T.}~\bibnamefont {{Sbarrato}}},
  \bibinfo {author} {\bibfnamefont {F.}~\bibnamefont {{Shankar}}},\ and\
  \bibinfo {author} {\bibfnamefont {M.}~\bibnamefont {{Colpi}}},\ }\href
  {https://doi.org/10.1093/mnras/stx2239} {\bibfield  {journal} {\bibinfo
  {journal} {\mnras}\ }\textbf {\bibinfo {volume} {472}},\ \bibinfo {pages}
  {4051} (\bibinfo {year} {2017})},\ \Eprint {https://arxiv.org/abs/1612.01580}
  {arXiv:1612.01580 [astro-ph.HE]} \BibitemShut {NoStop}%
\bibitem [{\citenamefont {{Runnoe}}\ \emph {et~al.}(2012)\citenamefont
  {{Runnoe}}, \citenamefont {{Brotherton}},\ and\ \citenamefont
  {{Shang}}}]{Runnoe12}%
  \BibitemOpen
  \bibfield  {author} {\bibinfo {author} {\bibfnamefont {J.~C.}\ \bibnamefont
  {{Runnoe}}}, \bibinfo {author} {\bibfnamefont {M.~S.}\ \bibnamefont
  {{Brotherton}}},\ and\ \bibinfo {author} {\bibfnamefont {Z.}~\bibnamefont
  {{Shang}}},\ }\href {https://doi.org/10.1111/j.1365-2966.2012.20620.x}
  {\bibfield  {journal} {\bibinfo  {journal} {\mnras}\ }\textbf {\bibinfo
  {volume} {422}},\ \bibinfo {pages} {478} (\bibinfo {year} {2012})},\ \Eprint
  {https://arxiv.org/abs/1201.5155} {arXiv:1201.5155 [astro-ph.CO]}
  \BibitemShut {NoStop}%
\bibitem [{\citenamefont {{Planck Collaboration}}\ \emph
  {et~al.}(2016)\citenamefont {{Planck Collaboration}} \emph {et~al.}}]{Ade16}%
  \BibitemOpen
  \bibfield  {author} {\bibinfo {author} {\bibnamefont {{Planck
  Collaboration}}} \emph {et~al.},\ }\href
  {https://doi.org/10.1051/0004-6361/201525830} {\bibfield  {journal} {\bibinfo
   {journal} {\aap}\ }\textbf {\bibinfo {volume} {594}},\ \bibinfo {eid} {A13}
  (\bibinfo {year} {2016})},\ \Eprint {https://arxiv.org/abs/1502.01589}
  {arXiv:1502.01589 [astro-ph.CO]} \BibitemShut {NoStop}%
\bibitem [{\citenamefont {{Stern}}\ \emph {et~al.}(2018)\citenamefont
  {{Stern}}, \citenamefont {{McKernan}}, \citenamefont {{Graham}} \emph
  {et~al.}}]{Stern18}%
  \BibitemOpen
  \bibfield  {author} {\bibinfo {author} {\bibfnamefont {D.}~\bibnamefont
  {{Stern}}}, \bibinfo {author} {\bibfnamefont {B.}~\bibnamefont {{McKernan}}},
  \bibinfo {author} {\bibfnamefont {M.~J.}\ \bibnamefont {{Graham}}}, \emph
  {et~al.},\ }\href {https://doi.org/10.3847/1538-4357/aac726} {\bibfield
  {journal} {\bibinfo  {journal} {\apj}\ }\textbf {\bibinfo {volume} {864}},\
  \bibinfo {eid} {27} (\bibinfo {year} {2018})},\ \Eprint
  {https://arxiv.org/abs/1805.06920} {arXiv:1805.06920 [astro-ph.GA]}
  \BibitemShut {NoStop}%
\bibitem [{\citenamefont {{Ross}}\ \emph {et~al.}(2018)\citenamefont {{Ross}},
  \citenamefont {{Ford}}, \citenamefont {{Graham}} \emph {et~al.}}]{Ross18}%
  \BibitemOpen
  \bibfield  {author} {\bibinfo {author} {\bibfnamefont {N.~P.}\ \bibnamefont
  {{Ross}}}, \bibinfo {author} {\bibfnamefont {K.~E.~S.}\ \bibnamefont
  {{Ford}}}, \bibinfo {author} {\bibfnamefont {M.}~\bibnamefont {{Graham}}},
  \emph {et~al.},\ }\href {https://doi.org/10.1093/mnras/sty2002} {\bibfield
  {journal} {\bibinfo  {journal} {\mnras}\ }\textbf {\bibinfo {volume} {480}},\
  \bibinfo {pages} {4468} (\bibinfo {year} {2018})},\ \Eprint
  {https://arxiv.org/abs/1805.06921} {arXiv:1805.06921 [astro-ph.GA]}
  \BibitemShut {NoStop}%
\bibitem [{\citenamefont {{Assef}}\ \emph
  {et~al.}(2018{\natexlab{a}})\citenamefont {{Assef}}, \citenamefont {{Stern}},
  \citenamefont {{Noirot}}, \citenamefont {{Jun}}, \citenamefont {{Cutri}},\
  and\ \citenamefont {{Eisenhardt}}}]{Assef18}%
  \BibitemOpen
  \bibfield  {author} {\bibinfo {author} {\bibfnamefont {R.~J.}\ \bibnamefont
  {{Assef}}}, \bibinfo {author} {\bibfnamefont {D.}~\bibnamefont {{Stern}}},
  \bibinfo {author} {\bibfnamefont {G.}~\bibnamefont {{Noirot}}}, \bibinfo
  {author} {\bibfnamefont {H.~D.}\ \bibnamefont {{Jun}}}, \bibinfo {author}
  {\bibfnamefont {R.~M.}\ \bibnamefont {{Cutri}}},\ and\ \bibinfo {author}
  {\bibfnamefont {P.~R.~M.}\ \bibnamefont {{Eisenhardt}}},\ }\href
  {https://doi.org/10.3847/1538-4365/aaa00a} {\bibfield  {journal} {\bibinfo
  {journal} {ApJS}\ }\textbf {\bibinfo {volume} {234}},\ \bibinfo {eid} {23}
  (\bibinfo {year} {2018}{\natexlab{a}})},\ \Eprint
  {https://arxiv.org/abs/1706.09901} {arXiv:1706.09901 [astro-ph.GA]}
  \BibitemShut {NoStop}%
\bibitem [{\citenamefont {{Kelly}}\ \emph {et~al.}(2009)\citenamefont
  {{Kelly}}, \citenamefont {{Bechtold}},\ and\ \citenamefont
  {{Siemiginowska}}}]{Kelly09}%
  \BibitemOpen
  \bibfield  {author} {\bibinfo {author} {\bibfnamefont {B.~C.}\ \bibnamefont
  {{Kelly}}}, \bibinfo {author} {\bibfnamefont {J.}~\bibnamefont
  {{Bechtold}}},\ and\ \bibinfo {author} {\bibfnamefont {A.}~\bibnamefont
  {{Siemiginowska}}},\ }\href {https://doi.org/10.1088/0004-637X/698/1/895}
  {\bibfield  {journal} {\bibinfo  {journal} {\apj}\ }\textbf {\bibinfo
  {volume} {698}},\ \bibinfo {pages} {895} (\bibinfo {year} {2009})},\ \Eprint
  {https://arxiv.org/abs/0903.5315} {arXiv:0903.5315 [astro-ph.CO]}
  \BibitemShut {NoStop}%
\bibitem [{\citenamefont {{Moreno}}\ \emph {et~al.}(2019)\citenamefont
  {{Moreno}}, \citenamefont {{Vogeley}}, \citenamefont {{Richards}},\ and\
  \citenamefont {{Yu}}}]{Moreno19}%
  \BibitemOpen
  \bibfield  {author} {\bibinfo {author} {\bibfnamefont {J.}~\bibnamefont
  {{Moreno}}}, \bibinfo {author} {\bibfnamefont {M.~S.}\ \bibnamefont
  {{Vogeley}}}, \bibinfo {author} {\bibfnamefont {G.~T.}\ \bibnamefont
  {{Richards}}},\ and\ \bibinfo {author} {\bibfnamefont {W.}~\bibnamefont
  {{Yu}}},\ }\href {https://doi.org/10.1088/1538-3873/ab1597} {\bibfield
  {journal} {\bibinfo  {journal} {PASP}\ }\textbf {\bibinfo {volume} {131}},\
  \bibinfo {pages} {063001} (\bibinfo {year} {2019})},\ \Eprint
  {https://arxiv.org/abs/1811.00154} {arXiv:1811.00154 [astro-ph.IM]}
  \BibitemShut {NoStop}%
\bibitem [{\citenamefont {{Graham}}\ \emph {et~al.}(2017)\citenamefont
  {{Graham}}, \citenamefont {{Djorgovski}}, \citenamefont {{Drake}},
  \citenamefont {{Stern}}, \citenamefont {{Mahabal}}, \citenamefont
  {{Glikman}}, \citenamefont {{Larson}},\ and\ \citenamefont
  {{Christensen}}}]{Graham17}%
  \BibitemOpen
  \bibfield  {author} {\bibinfo {author} {\bibfnamefont {M.~J.}\ \bibnamefont
  {{Graham}}}, \bibinfo {author} {\bibfnamefont {S.~G.}\ \bibnamefont
  {{Djorgovski}}}, \bibinfo {author} {\bibfnamefont {A.~J.}\ \bibnamefont
  {{Drake}}}, \bibinfo {author} {\bibfnamefont {D.}~\bibnamefont {{Stern}}},
  \bibinfo {author} {\bibfnamefont {A.~A.}\ \bibnamefont {{Mahabal}}}, \bibinfo
  {author} {\bibfnamefont {E.}~\bibnamefont {{Glikman}}}, \bibinfo {author}
  {\bibfnamefont {S.}~\bibnamefont {{Larson}}},\ and\ \bibinfo {author}
  {\bibfnamefont {E.}~\bibnamefont {{Christensen}}},\ }\href
  {https://doi.org/10.1093/mnras/stx1456} {\bibfield  {journal} {\bibinfo
  {journal} {\mnras}\ }\textbf {\bibinfo {volume} {470}},\ \bibinfo {pages}
  {4112} (\bibinfo {year} {2017})},\ \Eprint {https://arxiv.org/abs/1706.03079}
  {arXiv:1706.03079 [astro-ph.GA]} \BibitemShut {NoStop}%
\bibitem [{\citenamefont {{Drake}}\ \emph {et~al.}(2009)\citenamefont
  {{Drake}}, \citenamefont {{Djorgovski}}, \citenamefont {{Mahabal}},
  \citenamefont {{Beshore}}, \citenamefont {{Larson}}, \citenamefont
  {{Graham}}, \citenamefont {{Williams}}, \citenamefont {{Christensen}},
  \citenamefont {{Catelan}}, \citenamefont {{Boattini}}, \citenamefont
  {{Gibbs}}, \citenamefont {{Hill}},\ and\ \citenamefont
  {{Kowalski}}}]{Drake09}%
  \BibitemOpen
  \bibfield  {author} {\bibinfo {author} {\bibfnamefont {A.~J.}\ \bibnamefont
  {{Drake}}}, \bibinfo {author} {\bibfnamefont {S.~G.}\ \bibnamefont
  {{Djorgovski}}}, \bibinfo {author} {\bibfnamefont {A.}~\bibnamefont
  {{Mahabal}}}, \bibinfo {author} {\bibfnamefont {E.}~\bibnamefont
  {{Beshore}}}, \bibinfo {author} {\bibfnamefont {S.}~\bibnamefont {{Larson}}},
  \bibinfo {author} {\bibfnamefont {M.~J.}\ \bibnamefont {{Graham}}}, \bibinfo
  {author} {\bibfnamefont {R.}~\bibnamefont {{Williams}}}, \bibinfo {author}
  {\bibfnamefont {E.}~\bibnamefont {{Christensen}}}, \bibinfo {author}
  {\bibfnamefont {M.}~\bibnamefont {{Catelan}}}, \bibinfo {author}
  {\bibfnamefont {A.}~\bibnamefont {{Boattini}}}, \bibinfo {author}
  {\bibfnamefont {A.}~\bibnamefont {{Gibbs}}}, \bibinfo {author} {\bibfnamefont
  {R.}~\bibnamefont {{Hill}}},\ and\ \bibinfo {author} {\bibfnamefont
  {R.}~\bibnamefont {{Kowalski}}},\ }\href
  {https://doi.org/10.1088/0004-637X/696/1/870} {\bibfield  {journal} {\bibinfo
   {journal} {\apj}\ }\textbf {\bibinfo {volume} {696}},\ \bibinfo {pages}
  {870} (\bibinfo {year} {2009})},\ \Eprint {https://arxiv.org/abs/0809.1394}
  {arXiv:0809.1394 [astro-ph]} \BibitemShut {NoStop}%
\bibitem [{\citenamefont {{Assef}}\ \emph
  {et~al.}(2018{\natexlab{b}})\citenamefont {{Assef}}, \citenamefont
  {{Prieto}}, \citenamefont {{Stern}}, \citenamefont {{Cutri}}, \citenamefont
  {{Eisenhardt}}, \citenamefont {{Graham}}, \citenamefont {{Jun}},
  \citenamefont {{Rest}}, \citenamefont {{Flewelling}}, \citenamefont
  {{Kaiser}}, \citenamefont {{Kudritzki}},\ and\ \citenamefont
  {{Waters}}}]{Assef18sn}%
  \BibitemOpen
  \bibfield  {author} {\bibinfo {author} {\bibfnamefont {R.~J.}\ \bibnamefont
  {{Assef}}}, \bibinfo {author} {\bibfnamefont {J.~L.}\ \bibnamefont
  {{Prieto}}}, \bibinfo {author} {\bibfnamefont {D.}~\bibnamefont {{Stern}}},
  \bibinfo {author} {\bibfnamefont {R.~M.}\ \bibnamefont {{Cutri}}}, \bibinfo
  {author} {\bibfnamefont {P.~R.~M.}\ \bibnamefont {{Eisenhardt}}}, \bibinfo
  {author} {\bibfnamefont {M.~J.}\ \bibnamefont {{Graham}}}, \bibinfo {author}
  {\bibfnamefont {H.~D.}\ \bibnamefont {{Jun}}}, \bibinfo {author}
  {\bibfnamefont {A.}~\bibnamefont {{Rest}}}, \bibinfo {author} {\bibfnamefont
  {H.~A.}\ \bibnamefont {{Flewelling}}}, \bibinfo {author} {\bibfnamefont
  {N.}~\bibnamefont {{Kaiser}}}, \bibinfo {author} {\bibfnamefont {R.~P.}\
  \bibnamefont {{Kudritzki}}},\ and\ \bibinfo {author} {\bibfnamefont
  {C.}~\bibnamefont {{Waters}}},\ }\href
  {https://doi.org/10.3847/1538-4357/aaddf7} {\bibfield  {journal} {\bibinfo
  {journal} {\apj}\ }\textbf {\bibinfo {volume} {866}},\ \bibinfo {eid} {26}
  (\bibinfo {year} {2018}{\natexlab{b}})},\ \Eprint
  {https://arxiv.org/abs/1807.07985} {arXiv:1807.07985 [astro-ph.GA]}
  \BibitemShut {NoStop}%
\bibitem [{\citenamefont {{Kasen}}\ and\ \citenamefont
  {{Bildsten}}(2010)}]{Kasen10}%
  \BibitemOpen
  \bibfield  {author} {\bibinfo {author} {\bibfnamefont {D.}~\bibnamefont
  {{Kasen}}}\ and\ \bibinfo {author} {\bibfnamefont {L.}~\bibnamefont
  {{Bildsten}}},\ }\href {https://doi.org/10.1088/0004-637X/717/1/245}
  {\bibfield  {journal} {\bibinfo  {journal} {\apj}\ }\textbf {\bibinfo
  {volume} {717}},\ \bibinfo {pages} {245} (\bibinfo {year} {2010})},\ \Eprint
  {https://arxiv.org/abs/0911.0680} {arXiv:0911.0680 [astro-ph.HE]}
  \BibitemShut {NoStop}%
\bibitem [{\citenamefont {{Foley}}\ \emph {et~al.}(2011)\citenamefont
  {{Foley}}, \citenamefont {{Sanders}},\ and\ \citenamefont
  {{Kirshner}}}]{Foley11}%
  \BibitemOpen
  \bibfield  {author} {\bibinfo {author} {\bibfnamefont {R.~J.}\ \bibnamefont
  {{Foley}}}, \bibinfo {author} {\bibfnamefont {N.~E.}\ \bibnamefont
  {{Sanders}}},\ and\ \bibinfo {author} {\bibfnamefont {R.~P.}\ \bibnamefont
  {{Kirshner}}},\ }\href {https://doi.org/10.1088/0004-637X/742/2/89}
  {\bibfield  {journal} {\bibinfo  {journal} {\apj}\ }\textbf {\bibinfo
  {volume} {742}},\ \bibinfo {eid} {89} (\bibinfo {year} {2011})},\ \Eprint
  {https://arxiv.org/abs/1107.3555} {arXiv:1107.3555 [astro-ph.CO]}
  \BibitemShut {NoStop}%
\bibitem [{\citenamefont {{Lawrence}}\ \emph {et~al.}(2016)\citenamefont
  {{Lawrence}}, \citenamefont {{Bruce}}, \citenamefont {{MacLeod}},
  \citenamefont {{Gezari}}, \citenamefont {{Elvis}}, \citenamefont {{Ward}},
  \citenamefont {{Smartt}}, \citenamefont {{Smith}}, \citenamefont {{Wright}},
  \citenamefont {{Fraser}}, \citenamefont {{Marshall}}, \citenamefont
  {{Kaiser}}, \citenamefont {{Burgett}}, \citenamefont {{Magnier}},
  \citenamefont {{Tonry}}, \citenamefont {{Chambers}}, \citenamefont
  {{Wainscoat}}, \citenamefont {{Waters}}, \citenamefont {{Price}},
  \citenamefont {{Metcalfe}}, \citenamefont {{Valenti}}, \citenamefont
  {{Kotak}}, \citenamefont {{Mead}}, \citenamefont {{Inserra}}, \citenamefont
  {{Chen}},\ and\ \citenamefont {{Soderberg}}}]{Lawrence16}%
  \BibitemOpen
  \bibfield  {author} {\bibinfo {author} {\bibfnamefont {A.}~\bibnamefont
  {{Lawrence}}}, \bibinfo {author} {\bibfnamefont {A.~G.}\ \bibnamefont
  {{Bruce}}}, \bibinfo {author} {\bibfnamefont {C.}~\bibnamefont {{MacLeod}}},
  \bibinfo {author} {\bibfnamefont {S.}~\bibnamefont {{Gezari}}}, \bibinfo
  {author} {\bibfnamefont {M.}~\bibnamefont {{Elvis}}}, \bibinfo {author}
  {\bibfnamefont {M.}~\bibnamefont {{Ward}}}, \bibinfo {author} {\bibfnamefont
  {S.~J.}\ \bibnamefont {{Smartt}}}, \bibinfo {author} {\bibfnamefont {K.~W.}\
  \bibnamefont {{Smith}}}, \bibinfo {author} {\bibfnamefont {D.}~\bibnamefont
  {{Wright}}}, \bibinfo {author} {\bibfnamefont {M.}~\bibnamefont {{Fraser}}},
  \bibinfo {author} {\bibfnamefont {P.}~\bibnamefont {{Marshall}}}, \bibinfo
  {author} {\bibfnamefont {N.}~\bibnamefont {{Kaiser}}}, \bibinfo {author}
  {\bibfnamefont {W.}~\bibnamefont {{Burgett}}}, \bibinfo {author}
  {\bibfnamefont {E.}~\bibnamefont {{Magnier}}}, \bibinfo {author}
  {\bibfnamefont {J.}~\bibnamefont {{Tonry}}}, \bibinfo {author} {\bibfnamefont
  {K.}~\bibnamefont {{Chambers}}}, \bibinfo {author} {\bibfnamefont
  {R.}~\bibnamefont {{Wainscoat}}}, \bibinfo {author} {\bibfnamefont
  {C.}~\bibnamefont {{Waters}}}, \bibinfo {author} {\bibfnamefont
  {P.}~\bibnamefont {{Price}}}, \bibinfo {author} {\bibfnamefont
  {N.}~\bibnamefont {{Metcalfe}}}, \bibinfo {author} {\bibfnamefont
  {S.}~\bibnamefont {{Valenti}}}, \bibinfo {author} {\bibfnamefont
  {R.}~\bibnamefont {{Kotak}}}, \bibinfo {author} {\bibfnamefont
  {A.}~\bibnamefont {{Mead}}}, \bibinfo {author} {\bibfnamefont
  {C.}~\bibnamefont {{Inserra}}}, \bibinfo {author} {\bibfnamefont {T.~W.}\
  \bibnamefont {{Chen}}},\ and\ \bibinfo {author} {\bibfnamefont
  {A.}~\bibnamefont {{Soderberg}}},\ }\href
  {https://doi.org/10.1093/mnras/stw1963} {\bibfield  {journal} {\bibinfo
  {journal} {\mnras}\ }\textbf {\bibinfo {volume} {463}},\ \bibinfo {pages}
  {296} (\bibinfo {year} {2016})},\ \Eprint {https://arxiv.org/abs/1605.07842}
  {arXiv:1605.07842 [astro-ph.HE]} \BibitemShut {NoStop}%
\bibitem [{\citenamefont {{Rees}}(1988)}]{rees88}%
  \BibitemOpen
  \bibfield  {author} {\bibinfo {author} {\bibfnamefont {M.~J.}\ \bibnamefont
  {{Rees}}},\ }\href {https://doi.org/10.1038/333523a0} {\bibfield  {journal}
  {\bibinfo  {journal} {\nat}\ }\textbf {\bibinfo {volume} {333}},\ \bibinfo
  {pages} {523} (\bibinfo {year} {1988})}\BibitemShut {NoStop}%
\bibitem [{\citenamefont {{McKernan}}\ \emph
  {et~al.}(2020{\natexlab{b}})\citenamefont {{McKernan}}, \citenamefont
  {{Ford}},\ and\ \citenamefont {{O'Shaughnessy}}}]{McK20}%
  \BibitemOpen
  \bibfield  {author} {\bibinfo {author} {\bibfnamefont {B.}~\bibnamefont
  {{McKernan}}}, \bibinfo {author} {\bibfnamefont {K.~E.~S.}\ \bibnamefont
  {{Ford}}},\ and\ \bibinfo {author} {\bibfnamefont {R.}~\bibnamefont
  {{O'Shaughnessy}}},\ }\href@noop {} {\bibfield  {journal} {\bibinfo
  {journal} {arXiv e-prints}\ ,\ \bibinfo {eid} {arXiv:2002.00046}} (\bibinfo
  {year} {2020}{\natexlab{b}})},\ \Eprint {https://arxiv.org/abs/2002.00046}
  {arXiv:2002.00046 [astro-ph.HE]} \BibitemShut {NoStop}%
\bibitem [{\citenamefont {{Cannizzaro}}\ \emph {et~al.}(2020)\citenamefont
  {{Cannizzaro}}, \citenamefont {{Fraser}}, \citenamefont {{Jonker}} \emph
  {et~al.}}]{Cannizzaro20}%
  \BibitemOpen
  \bibfield  {author} {\bibinfo {author} {\bibfnamefont {G.}~\bibnamefont
  {{Cannizzaro}}}, \bibinfo {author} {\bibfnamefont {M.}~\bibnamefont
  {{Fraser}}}, \bibinfo {author} {\bibfnamefont {P.~G.}\ \bibnamefont
  {{Jonker}}}, \emph {et~al.},\ }\href {https://doi.org/10.1093/mnras/staa186}
  {\bibfield  {journal} {\bibinfo  {journal} {\mnras}\ ,\ \bibinfo {pages}
  {180}} (\bibinfo {year} {2020})},\ \Eprint {https://arxiv.org/abs/2001.07446}
  {arXiv:2001.07446 [astro-ph.HE]} \BibitemShut {NoStop}%
\bibitem [{\citenamefont {{Rosswog}}\ \emph {et~al.}(2009)\citenamefont
  {{Rosswog}}, \citenamefont {{Ramirez-Ruiz}},\ and\ \citenamefont
  {{Hix}}}]{rosswog09}%
  \BibitemOpen
  \bibfield  {author} {\bibinfo {author} {\bibfnamefont {S.}~\bibnamefont
  {{Rosswog}}}, \bibinfo {author} {\bibfnamefont {E.}~\bibnamefont
  {{Ramirez-Ruiz}}},\ and\ \bibinfo {author} {\bibfnamefont {W.~R.}\
  \bibnamefont {{Hix}}},\ }\href {https://doi.org/10.1088/0004-637X/695/1/404}
  {\bibfield  {journal} {\bibinfo  {journal} {\apj}\ }\textbf {\bibinfo
  {volume} {695}},\ \bibinfo {pages} {404} (\bibinfo {year} {2009})},\ \Eprint
  {https://arxiv.org/abs/0808.2143} {arXiv:0808.2143 [astro-ph]} \BibitemShut
  {NoStop}%
\bibitem [{\citenamefont {{Berry}}\ \emph {et~al.}(2015)\citenamefont
  {{Berry}}, \citenamefont {{Mandel}}, \citenamefont {{Middleton}} \emph
  {et~al.}}]{Berry15}%
  \BibitemOpen
  \bibfield  {author} {\bibinfo {author} {\bibfnamefont {C.~P.~L.}\
  \bibnamefont {{Berry}}}, \bibinfo {author} {\bibfnamefont {I.}~\bibnamefont
  {{Mandel}}}, \bibinfo {author} {\bibfnamefont {H.}~\bibnamefont
  {{Middleton}}}, \emph {et~al.},\ }\href
  {https://doi.org/10.1088/0004-637X/804/2/114} {\bibfield  {journal} {\bibinfo
   {journal} {\apj}\ }\textbf {\bibinfo {volume} {804}},\ \bibinfo {eid} {114}
  (\bibinfo {year} {2015})},\ \Eprint {https://arxiv.org/abs/1411.6934}
  {arXiv:1411.6934 [astro-ph.HE]} \BibitemShut {NoStop}%
\bibitem [{\citenamefont {{Finn}}\ and\ \citenamefont
  {{Chernoff}}(1993)}]{FinnChernoff93}%
  \BibitemOpen
  \bibfield  {author} {\bibinfo {author} {\bibfnamefont {L.~S.}\ \bibnamefont
  {{Finn}}}\ and\ \bibinfo {author} {\bibfnamefont {D.~F.}\ \bibnamefont
  {{Chernoff}}},\ }\href {https://doi.org/10.1103/PhysRevD.47.2198} {\bibfield
  {journal} {\bibinfo  {journal} {\prd}\ }\textbf {\bibinfo {volume} {47}},\
  \bibinfo {pages} {2198} (\bibinfo {year} {1993})},\ \Eprint
  {https://arxiv.org/abs/gr-qc/9301003} {arXiv:gr-qc/9301003 [gr-qc]}
  \BibitemShut {NoStop}%
\bibitem [{\citenamefont {{Abbott}}(2020)}]{2020arXiv200408342T}%
  \BibitemOpen
  \bibfield  {author} {\bibinfo {author} {\bibfnamefont {e.~a.}\ \bibnamefont
  {{Abbott}}, \bibfnamefont {R.}},\ }\href@noop {} {\bibfield  {journal}
  {\bibinfo  {journal} {arXiv e-prints}\ ,\ \bibinfo {eid} {arXiv:2004.08342}}
  (\bibinfo {year} {2020})},\ \Eprint {https://arxiv.org/abs/2004.08342}
  {arXiv:2004.08342 [astro-ph.HE]} \BibitemShut {NoStop}%
\bibitem [{\citenamefont {{Campanelli}}\ \emph {et~al.}(2007)\citenamefont
  {{Campanelli}}, \citenamefont {{Lousto}}, \citenamefont {{Zlochower}},\ and\
  \citenamefont {{Merritt}}}]{Campanelli07}%
  \BibitemOpen
  \bibfield  {author} {\bibinfo {author} {\bibfnamefont {M.}~\bibnamefont
  {{Campanelli}}}, \bibinfo {author} {\bibfnamefont {C.}~\bibnamefont
  {{Lousto}}}, \bibinfo {author} {\bibfnamefont {Y.}~\bibnamefont
  {{Zlochower}}},\ and\ \bibinfo {author} {\bibfnamefont {D.}~\bibnamefont
  {{Merritt}}},\ }\href {https://doi.org/10.1086/516712} {\bibfield  {journal}
  {\bibinfo  {journal} {\apjl}\ }\textbf {\bibinfo {volume} {659}},\ \bibinfo
  {pages} {L5} (\bibinfo {year} {2007})},\ \Eprint
  {https://arxiv.org/abs/gr-qc/0701164} {arXiv:gr-qc/0701164 [gr-qc]}
  \BibitemShut {NoStop}%
\bibitem [{\citenamefont {{Gonz{\'a}lez}}\ \emph {et~al.}(2007)\citenamefont
  {{Gonz{\'a}lez}}, \citenamefont {{Hannam}}, \citenamefont {{Sperhake}},
  \citenamefont {{Br{\"u}gmann}},\ and\ \citenamefont {{Husa}}}]{Gonzalez07}%
  \BibitemOpen
  \bibfield  {author} {\bibinfo {author} {\bibfnamefont {J.~A.}\ \bibnamefont
  {{Gonz{\'a}lez}}}, \bibinfo {author} {\bibfnamefont {M.}~\bibnamefont
  {{Hannam}}}, \bibinfo {author} {\bibfnamefont {U.}~\bibnamefont
  {{Sperhake}}}, \bibinfo {author} {\bibfnamefont {B.}~\bibnamefont
  {{Br{\"u}gmann}}},\ and\ \bibinfo {author} {\bibfnamefont {S.}~\bibnamefont
  {{Husa}}},\ }\href {https://doi.org/10.1103/PhysRevLett.98.231101} {\bibfield
   {journal} {\bibinfo  {journal} {\prl}\ }\textbf {\bibinfo {volume} {98}},\
  \bibinfo {eid} {231101} (\bibinfo {year} {2007})},\ \Eprint
  {https://arxiv.org/abs/gr-qc/0702052} {arXiv:gr-qc/0702052 [gr-qc]}
  \BibitemShut {NoStop}%
\bibitem [{\citenamefont {{Ostriker}}(1999)}]{Ostriker99}%
  \BibitemOpen
  \bibfield  {author} {\bibinfo {author} {\bibfnamefont {E.~C.}\ \bibnamefont
  {{Ostriker}}},\ }\href {https://doi.org/10.1086/306858} {\bibfield  {journal}
  {\bibinfo  {journal} {\apj}\ }\textbf {\bibinfo {volume} {513}},\ \bibinfo
  {pages} {252} (\bibinfo {year} {1999})},\ \Eprint
  {https://arxiv.org/abs/astro-ph/9810324} {arXiv:astro-ph/9810324 [astro-ph]}
  \BibitemShut {NoStop}%
\bibitem [{\citenamefont {{Antoni}}\ \emph {et~al.}(2019)\citenamefont
  {{Antoni}}, \citenamefont {{MacLeod}},\ and\ \citenamefont
  {{Ramirez-Ruiz}}}]{Antoni19}%
  \BibitemOpen
  \bibfield  {author} {\bibinfo {author} {\bibfnamefont {A.}~\bibnamefont
  {{Antoni}}}, \bibinfo {author} {\bibfnamefont {M.}~\bibnamefont
  {{MacLeod}}},\ and\ \bibinfo {author} {\bibfnamefont {E.}~\bibnamefont
  {{Ramirez-Ruiz}}},\ }\bibfield  {journal} {\bibinfo  {journal} {\apj}\
  }\textbf {\bibinfo {volume} {884}},\ \href
  {https://doi.org/10.3847/1538-4357/ab3466} {10.3847/1538-4357/ab3466}
  (\bibinfo {year} {2019}),\ \Eprint {https://arxiv.org/abs/1901.07572v1}
  {arXiv:1901.07572v1 [astro-ph.HE]} \BibitemShut {NoStop}%
\bibitem [{\citenamefont {{Tejeda}}\ and\ \citenamefont
  {{Aguayo-Ortiz}}(2019)}]{Tejeda19}%
  \BibitemOpen
  \bibfield  {author} {\bibinfo {author} {\bibfnamefont {E.}~\bibnamefont
  {{Tejeda}}}\ and\ \bibinfo {author} {\bibfnamefont {A.}~\bibnamefont
  {{Aguayo-Ortiz}}},\ }\href {https://doi.org/10.1093/mnras/stz1513} {\bibfield
   {journal} {\bibinfo  {journal} {\mnras}\ }\textbf {\bibinfo {volume}
  {487}},\ \bibinfo {pages} {3607} (\bibinfo {year} {2019})},\ \Eprint
  {https://arxiv.org/abs/1906.04923} {arXiv:1906.04923 [astro-ph.HE]}
  \BibitemShut {NoStop}%
\bibitem [{\citenamefont {{Zlochower}}\ and\ \citenamefont
  {{Lousto}}(2015)}]{Zloch15}%
  \BibitemOpen
  \bibfield  {author} {\bibinfo {author} {\bibfnamefont {Y.}~\bibnamefont
  {{Zlochower}}}\ and\ \bibinfo {author} {\bibfnamefont {C.~O.}\ \bibnamefont
  {{Lousto}}},\ }\href {https://doi.org/10.1103/PhysRevD.92.024022} {\bibfield
  {journal} {\bibinfo  {journal} {\prd}\ }\textbf {\bibinfo {volume} {92}},\
  \bibinfo {eid} {024022} (\bibinfo {year} {2015})},\ \Eprint
  {https://arxiv.org/abs/1503.07536} {arXiv:1503.07536 [gr-qc]} \BibitemShut
  {NoStop}%
\bibitem [{\citenamefont {{Healy}}\ and\ \citenamefont
  {{Lousto}}(2017)}]{Healy17}%
  \BibitemOpen
  \bibfield  {author} {\bibinfo {author} {\bibfnamefont {J.}~\bibnamefont
  {{Healy}}}\ and\ \bibinfo {author} {\bibfnamefont {C.~O.}\ \bibnamefont
  {{Lousto}}},\ }\href {https://doi.org/10.1103/PhysRevD.95.024037} {\bibfield
  {journal} {\bibinfo  {journal} {\prd}\ }\textbf {\bibinfo {volume} {95}},\
  \bibinfo {eid} {024037} (\bibinfo {year} {2017})},\ \Eprint
  {https://arxiv.org/abs/1610.09713} {arXiv:1610.09713 [gr-qc]} \BibitemShut
  {NoStop}%
\bibitem [{\citenamefont {{Sirko}}\ and\ \citenamefont
  {{Goodman}}(2003)}]{Sirko03}%
  \BibitemOpen
  \bibfield  {author} {\bibinfo {author} {\bibfnamefont {E.}~\bibnamefont
  {{Sirko}}}\ and\ \bibinfo {author} {\bibfnamefont {J.}~\bibnamefont
  {{Goodman}}},\ }\href {https://doi.org/10.1046/j.1365-8711.2003.06431.x}
  {\bibfield  {journal} {\bibinfo  {journal} {\mnras}\ }\textbf {\bibinfo
  {volume} {341}},\ \bibinfo {pages} {501} (\bibinfo {year} {2003})},\ \Eprint
  {https://arxiv.org/abs/astro-ph/0209469} {arXiv:astro-ph/0209469 [astro-ph]}
  \BibitemShut {NoStop}%
\bibitem [{\citenamefont {{Thompson}}\ \emph {et~al.}(2005)\citenamefont
  {{Thompson}}, \citenamefont {{Quataert}},\ and\ \citenamefont
  {{Murray}}}]{TQM05}%
  \BibitemOpen
  \bibfield  {author} {\bibinfo {author} {\bibfnamefont {T.~A.}\ \bibnamefont
  {{Thompson}}}, \bibinfo {author} {\bibfnamefont {E.}~\bibnamefont
  {{Quataert}}},\ and\ \bibinfo {author} {\bibfnamefont {N.}~\bibnamefont
  {{Murray}}},\ }\href {https://doi.org/10.1086/431923} {\bibfield  {journal}
  {\bibinfo  {journal} {\apj}\ }\textbf {\bibinfo {volume} {630}},\ \bibinfo
  {pages} {167} (\bibinfo {year} {2005})},\ \Eprint
  {https://arxiv.org/abs/astro-ph/0503027} {arXiv:astro-ph/0503027 [astro-ph]}
  \BibitemShut {NoStop}%
\bibitem [{\citenamefont {{Berti}}\ and\ \citenamefont
  {{Volonteri}}(2008)}]{BertiVolonteri08}%
  \BibitemOpen
  \bibfield  {author} {\bibinfo {author} {\bibfnamefont {E.}~\bibnamefont
  {{Berti}}}\ and\ \bibinfo {author} {\bibfnamefont {M.}~\bibnamefont
  {{Volonteri}}},\ }\href {https://doi.org/10.1086/590379} {\bibfield
  {journal} {\bibinfo  {journal} {\apj}\ }\textbf {\bibinfo {volume} {684}},\
  \bibinfo {pages} {822} (\bibinfo {year} {2008})},\ \Eprint
  {https://arxiv.org/abs/0802.0025} {arXiv:0802.0025 [astro-ph]} \BibitemShut
  {NoStop}%
\bibitem [{\citenamefont {{Gerosa}}\ and\ \citenamefont
  {{Berti}}(2017)}]{GerosaBerti17}%
  \BibitemOpen
  \bibfield  {author} {\bibinfo {author} {\bibfnamefont {D.}~\bibnamefont
  {{Gerosa}}}\ and\ \bibinfo {author} {\bibfnamefont {E.}~\bibnamefont
  {{Berti}}},\ }\href {https://doi.org/10.1103/PhysRevD.95.124046} {\bibfield
  {journal} {\bibinfo  {journal} {\prd}\ }\textbf {\bibinfo {volume} {95}},\
  \bibinfo {eid} {124046} (\bibinfo {year} {2017})},\ \Eprint
  {https://arxiv.org/abs/1703.06223} {arXiv:1703.06223 [gr-qc]} \BibitemShut
  {NoStop}%
\bibitem [{\citenamefont {{Fishbach}}\ and\ \citenamefont
  {{Holz}}(2017)}]{Fishbach17}%
  \BibitemOpen
  \bibfield  {author} {\bibinfo {author} {\bibfnamefont {M.}~\bibnamefont
  {{Fishbach}}}\ and\ \bibinfo {author} {\bibfnamefont {D.~E.}\ \bibnamefont
  {{Holz}}},\ }\href {https://doi.org/10.3847/2041-8213/aa9bf6} {\bibfield
  {journal} {\bibinfo  {journal} {\apjl}\ }\textbf {\bibinfo {volume} {851}},\
  \bibinfo {eid} {L25} (\bibinfo {year} {2017})},\ \Eprint
  {https://arxiv.org/abs/1709.08584} {arXiv:1709.08584 [astro-ph.HE]}
  \BibitemShut {NoStop}%
\bibitem [{\citenamefont {{Gerosa}}\ and\ \citenamefont
  {{Moore}}(2016)}]{Gerosa16}%
  \BibitemOpen
  \bibfield  {author} {\bibinfo {author} {\bibfnamefont {D.}~\bibnamefont
  {{Gerosa}}}\ and\ \bibinfo {author} {\bibfnamefont {C.~J.}\ \bibnamefont
  {{Moore}}},\ }\href {https://doi.org/10.1103/PhysRevLett.117.011101}
  {\bibfield  {journal} {\bibinfo  {journal} {\prl}\ }\textbf {\bibinfo
  {volume} {117}},\ \bibinfo {eid} {011101} (\bibinfo {year} {2016})},\ \Eprint
  {https://arxiv.org/abs/1606.04226} {arXiv:1606.04226 [gr-qc]} \BibitemShut
  {NoStop}%
\bibitem [{\citenamefont {{Varma}}\ \emph {et~al.}(2020)\citenamefont
  {{Varma}}, \citenamefont {{Isi}},\ and\ \citenamefont
  {{Biscoveanu}}}]{Varma20}%
  \BibitemOpen
  \bibfield  {author} {\bibinfo {author} {\bibfnamefont {V.}~\bibnamefont
  {{Varma}}}, \bibinfo {author} {\bibfnamefont {M.}~\bibnamefont {{Isi}}},\
  and\ \bibinfo {author} {\bibfnamefont {S.}~\bibnamefont {{Biscoveanu}}},\
  }\href {https://doi.org/10.1103/PhysRevLett.124.101104} {\bibfield  {journal}
  {\bibinfo  {journal} {\prl}\ }\textbf {\bibinfo {volume} {124}},\ \bibinfo
  {eid} {101104} (\bibinfo {year} {2020})},\ \Eprint
  {https://arxiv.org/abs/2002.00296} {arXiv:2002.00296 [gr-qc]} \BibitemShut
  {NoStop}%
\end{thebibliography}%

\end{document}